\documentclass[12pt]{article}
\usepackage[utf8]{inputenc}
\usepackage{amsmath}
\usepackage{amssymb}
\usepackage{times}
\usepackage{txfonts}
\usepackage[mathlines,displaymath]{lineno}
\DeclareMathAlphabet{\mathbold}{OML}{txr}{b}{it}
\usepackage[font={small}]{caption}

\usepackage{multirow}
\usepackage{units}
\usepackage{epsfig}
\usepackage{hhline}
\usepackage{dsfont}
\usepackage{color}
\usepackage{pdflscape}
\usepackage{cite} 

\usepackage{longtable}
\usepackage{xspace}
\usepackage{bm} 
\usepackage{acronym}
\usepackage{dcolumn}
\newcolumntype{L}{D{.}{.}{2,5}}

\usepackage{graphicx}
\usepackage{rotating}
\usepackage{paralist} 

\usepackage{hyperref} 
\usepackage{xcolor}
\hypersetup{colorlinks=true, urlcolor=blue}
\usepackage{cite} 
\hypersetup{
  colorlinks,
  citecolor=blue,
  linkcolor=red,
  urlcolor=blue
  }
  
\renewcommand{\arraystretch}{1.25} 
\newlength{\dinwidth}
\newlength{\dinmargin}
\usepackage{parskip}
\begin{document}  

\newcommand{\delrel}{\ensuremath{\delta}}
\newcommand{\delabs}{\ensuremath{\Delta}}
\newcommand{\dabs}[2][1]{\ensuremath{\delabs^{{\rm{#1}}}_{{#2}}}}
\newcommand{\drel}[2][1]{\ensuremath{\delrel^{{\rm{#1}}}_{{#2}}}}
\newcommand{\Nsys}{{N_{\rm sys}}}
\newcommand{\stat}{{\rm stat}}
\newcommand{\sys}{{\rm sys}}
\newcommand{\TODO}{\color{red}TODO\xspace}
\newcommand{\AS}{\color{red} AS:\xspace}

\newcommand{\muf}{\ensuremath{\mu_{f}}\xspace}
\newcommand{\mur}{\ensuremath{\mu_{r}}\xspace}
\newcommand{\as}{\ensuremath{\alpha_s}\xspace}
\newcommand{\asmz}{\ensuremath{\alpha_s(M_Z)}\xspace}
\newcommand{\asmur}{\ensuremath{\alpha_s(\mur)}\xspace}
\newcommand{\aem}{\ensuremath{\alpha_{\mathrm{em}}}\xspace}
\newcommand{\Lumi}{\ensuremath{\mathcal{L}}}
\newcommand{\pb}{\rm pb}
\newcommand{\invpb}{\ensuremath{\rm{pb}^{-1}}}
\newcommand{\PDF}{\ensuremath{{\rm PDF}}\xspace}
\renewcommand{\deg}{\ensuremath{^\circ}\xspace}
\newcommand{\unitmatrix}{1\!\!1}
\newcommand{\fC}{\ensuremath{f^{\rm C}}\xspace}
\newcommand{\fU}{\ensuremath{f^{\rm U}}\xspace}
\newcommand{\bas}{\boldsymbol{{\alpha_s}}} 
\newcommand{\basmz}{\boldsymbol{\alpha_s(M_Z)}} 
\newcommand{\bmur}{\boldsymbol{\mu_{r}}}

\newcommand{\chisq}{\ensuremath{\chi^{2}}}
\newcommand{\chisqA}{\ensuremath{\chi_{\rm A}^{2}}}
\newcommand{\chisqL}{\ensuremath{\chi_{\rm L}^{2}}}
\newcommand{\ndf}{\ensuremath{n_{\rm dof}}}
\newcommand{\A}{\ensuremath{\bm{A}}}
\newcommand{\M}{\ensuremath{\bm{M}}}
\newcommand{\V}{\ensuremath{\bm{V}}}
\newcommand{\B}{\ensuremath{\bm{B}}}
\newcommand{\J}{\ensuremath{\bm{J}}}
\newcommand{\N}{\ensuremath{\bm{N}}}
\newcommand{\LL}{\ensuremath{\bm{L}}}

\newcommand{\femjet}{\ensuremath{f_{\mathrm{em,jet}}}\xspace}
\newcommand{\femjetgen}{\ensuremath{f_{\mathrm{em,jet}}^{\mathrm{gen}}}\xspace}
\newcommand{\femjetrec}{\ensuremath{f_{\mathrm{em,jet}}^{\mathrm{rec}}}\xspace}
\newcommand{\Pem}{\ensuremath{P_{\mathrm{em}}}\xspace}
\newcommand{\Eem}{\ensuremath{E_{\mathrm{em}}}\xspace}
\newcommand{\Ehad}{\ensuremath{E_{\mathrm{had}}}\xspace}
\newcommand{\Ptbal}{\ensuremath{P_{\mathrm{T}}}--balance\ }
\newcommand{\PTbal}{\ensuremath{P_{\mathrm{T,bal}}}\xspace}
\newcommand{\Ptgen}{\ensuremath{P_{\mathrm{T}}^\mathrm{gen}}\xspace}
\newcommand{\Ptda}{\ensuremath{P_{\mathrm{T}}^{\mathrm{da}}}\xspace}
\newcommand{\thetajet}{\ensuremath{\theta_{\mathrm{jet}}}\xspace}
\newcommand{\etajet}{\ensuremath{\eta_{\mathrm{jet}}}\xspace}
\newcommand{\Ejet}{\ensuremath{E_{\mathrm{jet}}}\xspace}
\newcommand{\Ejetda}{\ensuremath{E^{\mathrm{da}}_{\mathrm{jet}}}\xspace}
\newcommand{\relres}{\ensuremath{\sigma(\Ejet)/\Ejet}\xspace}

\newcommand{\Empz}{\ensuremath{E-p_z}\xspace}
\newcommand{\gammah}{\ensuremath{\gamma_{\mathrm{h}}}\xspace}
\newcommand{\Pth}{\ensuremath{P_{\mathrm{T}}^{\mathrm{h}}}\xspace}
\newcommand{\Pzh}{\ensuremath{p_z^{\mathrm{h}}}\xspace}
\newcommand{\h}{\ensuremath{\mathrm{h}}}

\newcommand{\e}{\ensuremath{\mathrm{e}}}
\newcommand{\Ee}{\ensuremath{E_\mathrm{e}^\prime}\xspace}
\newcommand{\Pte}{\ensuremath{P_{\mathrm{T}}^{\mathrm{e}}}\xspace}
\newcommand{\thetae}{\ensuremath{\theta_\mathrm{e^\prime}}\xspace}
\newcommand{\phie}{\ensuremath{\phi_\mathrm{e}}\xspace}
\newcommand{\Eda}{\ensuremath{E^{\mathrm{da}}}\xspace}

\newcommand{\Ebeam}{\ensuremath{E_{\mathrm{e}}}\xspace}
\newcommand{\Pbeam}{\ensuremath{E_{\mathrm{p}}}\xspace}

\newcommand{\xbj}{\ensuremath{x}\xspace}
\newcommand{\Qsq}{\ensuremath{Q^2}\xspace}

\newcommand{\ep}{\ensuremath{e^+}\xspace}
\newcommand{\emm}{\ensuremath{e^-}\xspace}
\newcommand{\epm}{\ensuremath{e^\pm}\xspace}

\newcommand{\ptjet}{\ensuremath{P_{\rm T}^{\rm jet}}\xspace}
\newcommand{\ptjetmin}{\ensuremath{P_{\rm T,min}^{\rm jet}}\xspace}
\newcommand{\meanpt}{\ensuremath{\langle P_{\rm T} \rangle}}
\newcommand{\meanptdi}{\ensuremath{\langle P_{\mathrm{T}} \rangle_{2}}\xspace}
\newcommand{\meanpttri}{\ensuremath{\langle P_{\mathrm{T}} \rangle_{3}}\xspace}
\newcommand{\pt}{\ensuremath{P_{\rm T}}\xspace}
\newcommand{\Pt}{\ensuremath{P_{\mathrm{T}}}\xspace}
\newcommand{\Ptone}{\ensuremath{P_{\mathrm{T,1}}}\xspace}
\newcommand{\Pttwo}{\ensuremath{P_{\mathrm{T,2}}}\xspace}

\newcommand{\kt}{\ensuremath{{k_{\mathrm{T}}}}\xspace}
\newcommand{\antikt}{\ensuremath{{\mathrm{anti-}k_{\mathrm{T}}}}\xspace}
\newcommand{\bkt}{\ensuremath{\bm{k}_{\boldsymbol{\mathrm{T}}} }\xspace}
\newcommand{\bantikt}{\ensuremath{\boldsymbol{\mathrm{anti-}}\bm{k}_{\boldsymbol{\mathrm{T}}}}\xspace}

\newcommand{\etal}{{\it et al.}}
\newcommand{\Hone}{[H1 Collaboration]}

\newcommand{\Et}{\ensuremath{E_\mathrm{T}}\xspace}
\newcommand{\etalab}{\ensuremath{\eta^{\mathrm{jet}}_{\mathrm{lab}}}\xspace}
\newcommand{\ptlab}{\ensuremath{P^{\mathrm{jet}}_{\mathrm{T,lab}}}\xspace}
\newcommand{\Mjj}{\ensuremath{M_{\mathrm{12}}}\xspace}
\newcommand{\Mjjj}{\ensuremath{M_{\rm 123}}}
\newcommand{\xij}{\ensuremath{\xi}\xspace}
\newcommand{\xidi}{\ensuremath{\xi_2}\xspace}
\newcommand{\xitri}{\ensuremath{\xi_3}\xspace}

\newcommand{\ud}{\ensuremath{\mathrm{d}}\xspace}
\newcommand{\LO}{\ensuremath{\mathcal{O}(\alpha_s^0)}\xspave}
\newcommand{\Oa}{\ensuremath{\mathcal{O}(\alpha_s)}\xspace}
\newcommand{\Oaa}{\ensuremath{\mathcal{O}(\alpha_s^2)}\xspace}
\newcommand{\Oaaa}{\ensuremath{\mathcal{O}(\alpha_s^3)}\xspace}

\newcommand{\eq}{equation}
\newcommand{\fig}{figure}
\newcommand{\tab}{table}

\newcommand{\MeV}{\ensuremath{\mathrm{MeV}}\xspace}
\newcommand{\GeV}{\ensuremath{\mathrm{GeV}}\xspace}
\newcommand{\GeVsq}{\ensuremath{\mathrm{GeV}^2}\xspace}
\newcommand{\sw}{\ensuremath{{\rm sin}^2\theta_W}}
\newcommand{\dr}{\ensuremath{\Delta r}}
\newcommand{\gf}{\ensuremath{G_{\rm F}}}

\newcommand{\rhopW}[2][]{\ensuremath{\rho^{\prime}_{\text{CC}#2}}}
\newcommand{\rhop}[2][] {\ensuremath{\rho^{\prime}_{\text{NC}#2}}}
\newcommand{\kapp}[2][] {\ensuremath{\kappa^{\prime}_{\text{NC}#2}}}
\newcommand{\rhopu}{\rhop{, u}}
\newcommand{\kappu}{\kapp{, u}}
\newcommand{\rhopd}{\rhop{, d}}
\newcommand{\kappd}{\kapp{, d}}
\newcommand{\rhope}{\rhop{, e}}
\newcommand{\kappe}{\kapp{, e}}
\newcommand{\kapz}{\ensuremath{\kappa_{\text{NC}, f}}}
\newcommand{\rhoz}{\ensuremath{\rho_{\text{NC},f}}}
\newcommand{\Itf}{\ensuremath{I^3_{{\rm L},f}}}
\newcommand{\Itq}{\ensuremath{I^3_{{\rm L},q}}}

\newcommand{\fifteen}{\ensuremath{15\,000}}
\newcommand{\dd}{\mathrm{d}}
\newcommand{\Ord}{\ensuremath{\mathcal{O}}}

\newcommand{\ad} {\ensuremath{g_A^d}}
\newcommand{\vd} {\ensuremath{g_V^d}}
\newcommand{\au} {\ensuremath{g_A^u}}
\newcommand{\vu} {\ensuremath{g_V^u}}
\newcommand{\aq} {\ensuremath{g_A^q}}
\newcommand{\vq} {\ensuremath{g_V^q}}
\newcommand{\gae}{\ensuremath{g_A^e}}
\newcommand{\ve} {\ensuremath{g_V^e}}

\newcommand{\mt} {\ensuremath{m_t}}
\newcommand{\mW} {\ensuremath{m_W}}
\newcommand{\mWprop} {\ensuremath{m^{\rm prop}_W}}
\newcommand{\mWGfW} {\ensuremath{m^{(\gf,\mW)}_W}}
\newcommand{\mZ} {\ensuremath{m_Z}}
\newcommand{\mH} {\ensuremath{m_H}}

\def\Journal#1#2#3#4{{#1}~{\bf #2} (#3) #4}
\def\NPB{Nucl. Phys.~}
\def\PRL{Phys. Rev. Lett.~}
\def\EPJC{Eur. Phys. J.~}
\def\PLB{Phys. Lett.~}
\def\NIM{Nucl. Instrum. Meth.~}
\def\PRD{Phys. Rev.~}
\def\JHEP{JHEP~}
\def\PROC{Conf. Proc.~}
\def\CPC{Comp. Phys. Commun.~}

\begin{titlepage}

\noindent
\begin{flushleft}
{\tt DESY 18-080    \hfill    ISSN 0418-9833} \\
\end{flushleft}

\noindent
Date:      \ \ \ May 2018      \\
\noindent

\vspace{2cm}
\begin{center}
\begin{Large}

{\bf Determination of electroweak parameters in polarised deep-inelastic scattering at HERA}

\vspace{2cm}

H1 Collaboration and H.~Spiesberger (Mainz)

\end{Large}
\end{center}

\vspace{2cm}

\begin{abstract}
\noindent
The parameters of the electroweak theory are determined in
a combined electroweak and QCD analysis using all deep-inelastic
$e^+p$ and $e^-p$ neutral current and
charged current scattering cross sections published by the H1
Collaboration,
including data with longitudinally polarised lepton beams.
Various fits to Standard Model parameters in the on-shell scheme
are performed.
The mass of the $W$ boson is determined as
$\mW=80.520\pm 0.115\,\GeV$. The axial-vector and vector couplings 
of the light quarks to the $Z$ boson are also determined.
Both results improve the precision of previous H1 determinations based on HERA-I
data by about a factor of two. 
Possible scale dependence of the weak coupling parameters in both neutral and
charged current interactions beyond the Standard Model is also studied.
All results are found to be consistent
with the Standard Model expectations. 
\end{abstract}

\vspace{1.5cm}
\centering
Dedicated to the memory of our dear friend and colleague Violette Brisson

\vspace{1cm}

\begin{center} Submitted to EPJ C \end{center}

\end{titlepage}
\sloppy

\begin{flushleft}
V.~Andreev$^{19}$,             
A.~Baghdasaryan$^{31}$,        
K.~Begzsuren$^{28}$,           
A.~Belousov$^{19}$,            
A.~Bolz$^{12}$,                
V.~Boudry$^{22}$,              
G.~Brandt$^{40}$,              
V.~Brisson$^{21, \dagger}$,    
D.~Britzger$^{12}$,            
A.~Buniatyan$^{2}$
% ...


\bigskip{\it
% ...
$ ^{2}$ School of Physics and Astronomy, University of Birmingham,
          Birmingham, UK$^{ b}$ \\
% ...
$ ^{12}$ Physikalisches Institut, Universit\"at Heidelberg,
          Heidelberg, Germany$^{ a}$ \\
% ...
$ ^{19}$ Lebedev Physical Institute, Moscow, Russia \\
% ...
$ ^{21}$ LAL, Universit\'e Paris-Sud, CNRS/IN2P3, Orsay, France \\
 $ ^{22}$ LLR, Ecole Polytechnique, CNRS/IN2P3, Palaiseau, France \\
% ...
$ ^{28}$ Institute of Physics and Technology of the Mongolian
          Academy of Sciences, Ulaanbaatar, Mongolia \\
% ...
$ ^{31}$ Yerevan Physics Institute, Yerevan, Armenia \\
% ...
$ ^{40}$ II. Physikalisches Institut, Universit\"at G\"ottingen,
          G\"ottingen, Germany \\
% ...
$ ^{\dagger}$ Deceased \\

\bigskip
 $ ^a$ Supported by the Bundesministerium f\"ur Bildung und Forschung, FRG,
      under contract numbers 05H09GUF, 05H09VHC, 05H09VHF,  05H16PEA \\
 $ ^b$ Supported by the UK Science and Technology Facilities Council,
      and formerly by the UK Particle Physics and
      Astronomy Research Council \\
% ...
}

\end{flushleft}
\clearpage
%


\section{Introduction}
Since the discovery of weak neutral currents in 1973~\cite{Hasert:1973ff,Haidt:2015bgg},
the Glashow-Weinberg-Salam
model~\cite{Glashow:1961tr,Weinberg:1967tq,Weinberg:1971fb,Weinberg:1972tu,Salam:1964ry,Higgs:1964ia,Higgs:1964pj,Englert:1964et}
has been established as the theory of electroweak (EW) interactions and as
the core of the Standard Model (SM) of particle physics.
Already since these early times, deep-inelastic lepton-hadron
scattering (DIS) experiments with longitudinally polarised electron beams have provided indispensable
results~\cite{Prescott:1978tm,Prescott:1979dh} for its great
success.
Nowadays, EW theory has been tested in great detail at
lower scales with muon life-time measurements~\cite{Tishchenko:2012ie}
and neutrino scattering experiments~\cite{Fogli:1988tv,Blondel:1989ev,Allaby:1987vr,McFarland:1997wx,Zeller:2001hh}, 
with precision measurements at the $Z$ pole and at even higher 
scales~\cite{ALEPH:2005ab,Chatrchyan:2011ya,Schael:2013ita,Aaij:2015lka,Aad:2015uau,Aaltonen:2018dxj}.
The H1 Collaboration has performed first studies of weak interactions at
the HERA electron-proton collider in 1993:
the measurement of the total charged-current cross section
demonstrated for the first 
time the presence of the $W$-boson propagator~\cite{Ahmed:1994fa}.
DIS at HERA provides complementary testing ground for studying EW
processes at the EW energy scale in the space-like regime.  
The centre-of-mass energy at HERA 
nicely fills the gap between low-energy neutrino or muon 
experiments and high-energy collider experiments, and it
offers the possibility to study neutral and charged currents (NC and
CC) on equal footing.

The H1 experiment~\cite{Andrieu:1993kh,Abt:1996hi,Abt:1996xv,Appuhn:1996na}
at the HERA collider recorded collisions of
electrons and positrons of 27.6\,\GeV\ and unpolarised protons of up to
920\,\GeV\ during the HERA-I running period in the years 1992 to 2000,
and the HERA-II running period in the years 2003 to 2007.
These data provide a large set
of precise NC and CC cross section measurements.
They are an important input to study Quantum
Chromodynamics (QCD),  the theory of the strong force, and are
indispensable for exploring the structure of the proton.
Furthermore, at the HERA centre-of-mass energy of up to $\sqrt{s}=319\,\GeV$, EW effects such as $\gamma Z$ interference significantly contribute to the inclusive NC DIS cross
sections at high values of negative four-momentum transfers squared ($\Qsq$). 
The CC interactions are solely mediated by charged $W$ bosons.
This allows for a determination of EW parameters
from inclusive NC and CC DIS data at high \Qsq\ up to $50\,000\,\GeVsq$

At HERA, several determinations of the $W$-boson mass ($\mW$) have been
performed by the H1 and ZEUS experiments
based on different data samples collected during the HERA-I data
taking
period~\cite{Aid:1996yc,Breitweg:1999aa,Adloff:1999ah,Adloff:2000qj,Chekanov:2002zs}.
A first EW analysis
was performed using the complete HERA-I data collected by H1~\cite{Aktas:2005iv}, where
the weak neutral-current couplings of the light quarks to the $Z$ boson, the axial-vector
($g_A^{u/d}$) and vector ($g_V^{u/d}$) couplings, and \mW\
and the top-quark mass ($m_t$) were determined. 
Analyses using H1 data from HERA-I and HERA-II cross section measurements
together with ZEUS data have been reported by the ZEUS Collaboration~\cite{Abramowicz:2016ztw} and by I.~Abt et al.~\cite{Abt:2016zth}.

In the present analysis, the entire set of inclusive NC and CC DIS cross sections
measured by the H1 Collaboration during the HERA-I and HERA-II running
periods is exploited.
The studies thus benefit from the improved statistical precision of the
data samples, as compared to the previous
analysis~\cite{Aktas:2005iv}.
In addition, the longitudinal polarisation of the lepton beams in the
HERA-II running provides new sensitivity.

The EW parameters are determined together with the parameters of   
parton density functions (PDFs) of the proton in combined fits, thus
accounting for their correlated uncertainties.
The cross section predictions used in this analysis include next-to-next-to-leading order (NNLO)
 QCD corrections at the hadronic vertex and next-to-leading order
(NLO) EW corrections.
Within the SM framework the masses of the $W$ and $Z$
bosons and the couplings of the light quarks are determined.
Potential modifications from physics beyond the SM are explored.
EW parameters are tested in DIS at space-like four-momentum transfer.
Therefore, the studies presented here are complementary to
measurements of EW parameters at $e^+e^-$ or $pp$ colliders, which are performed in the
time-like regime for example at the $Z$ pole or at the $WW$
threshold.

\section{Theoretical framework}
\label{text:section:theory}
NC interactions in the process $e^\pm p\rightarrow e^\pm X$ are
mediated by a virtual photon $(\gamma)$ or $Z$ boson in the
$t$-channel, and the cross section is expressed in terms of
generalised structure functions $\tilde{F}_2^\pm$, $x\tilde{F}_3^\pm$
and $\tilde{F}_{\rm L}^\pm$ at EW leading order (LO)
as
\begin{equation}
  \frac{d^2\sigma^{\rm NC}(e^\pm p)}{dxd\Qsq} = \frac{2\pi\alpha^2}{xQ^4}\left[Y_+\tilde{F}_2^\pm(x,\Qsq) \mp Y_{-}  x\tilde{F}_3^\pm(x,\Qsq) - y^2 \tilde{F}_{\rm L}^\pm(x,\Qsq)\right]~,
  \label{eq:cs}
\end{equation}
where $\alpha$ is the fine structure constant and $x$ denotes the
Bjorken scaling variable (see e.g.~\cite{CooperSarkar:1998ug}).
The helicity dependence of the interaction is contained in the terms $Y_\pm = 1\pm(1-y)^2$ with $y$ being the inelasticity of the process.
The generalised structure functions can be separated into contributions
from pure $\gamma$- and $Z$-exchange and their interference~\cite{Klein:1983vs},
\begin{align}
  \tilde{F}_2^\pm
  &= F_2
  -(\ve\pm P_e\gae)\varkappa_ZF_2^{\gamma Z}
  +\left[(\ve\ve+\gae\gae)\pm2P_e\ve\gae\right]\varkappa_Z^2F_2^Z~,
  \\
  \tilde{F}_3^\pm
  &=~~
  -(\gae\pm P_e\ve)\varkappa_ZF_3^{\gamma Z}
  +\left[2\ve\gae\pm P_e(\ve\ve+\gae\gae)\right]\varkappa_Z^2F_3^Z~,
\end{align}
and similarly for $\tilde{F}_L$. The variables $g^e_A$ and $g^e_V$
stand for the axial-vector and vector couplings of the lepton $e^\pm$
to the $Z$ boson, respectively.
The degree of longitudinal polarisation of the incoming lepton is
denoted as $P_e$.
 
The $\Qsq$-dependent coefficient
$\varkappa_Z$ accounts for the $Z$-boson propagator,
\begin{equation}
  \varkappa_Z(\Qsq)
  = \frac{\Qsq}{\Qsq+m^2_Z} 
  \frac{1}{4\sw \cos^2\theta_W}
  = \frac{\Qsq}{\Qsq+m^2_Z} 
  \frac{\gf m_Z^2}{2\sqrt{2}\pi\alpha}~.
\end{equation}
It can be normalised using the 
weak mixing angle, $\sw=1-\mW^2 / \mZ^2$, i.e.\ using the $W$ and $Z$ boson
masses, \mW\ and \mZ, or the
Fermi coupling constant $\gf$, which is measured with high precision
in muon-decay experiments~\cite{Tishchenko:2012ie}.
The structure functions are related to linear combinations of the
quark and anti-quark momentum distributions, $xq$ and $x\bar{q}$.
For instance, the $F_2$ and $xF_3$ structure functions in
the naive quark-parton model, i.e.\ at LO in QCD, are:
\begin{align}
  \left[F_2,F_2^{\gamma Z},F_2^Z\right]
  &= x\sum_q\left[Q_q^2,2Q_q\vq,\vq\vq+\aq\aq \right]\{q+\bar{q}\}~,
  \label{eq:last1}
  \\
  x\left[F_3^{\gamma Z},F_3^Z\right]
  &= x\sum_q\left[2Q_q\aq,2\vq\aq\right]\{q-\bar{q}\}~.
  \label{eq:last2}
\end{align}
The axial-vector and vector couplings of the quarks $q$ to the $Z$ boson,
$g^q_A$ and $g^q_V$, depend on the electric charge, $Q_q$, in units of the
positron charge,
and on the third component of the weak-isospin of the quarks, \Itq.
In terms of $\sw$, they are given by the standard EW theory:
\begin{align}
  g_A^q &= \Itq 
  \label{eq:gA-LO} \,, \\
  g_V^q &= \Itq - 2 Q_q \sw
  \label{eq:gV-LO} \,.
\end{align}
The same formulae also apply to the lepton couplings $g^e_{A/V}$.

Universal higher-order corrections, to be discussed below, 
can be taken into account by introducing $\Qsq$-dependent form 
factors $\rho_{\text{NC}, q}$ and $\kappa_{\text{NC}, q}$ \cite{Olive:2016xmw}, replacing 
equations~\eqref{eq:gA-LO} and \eqref{eq:gV-LO} by 
\begin{align}
  g_A^q &= \sqrt{\rho_{\text{NC}, q}} \Itq 
  \label{eq:gA-NLO} \, , 
  \\
  g_V^q &= \sqrt{\rho_{\text{NC}, q}} \left( \Itq - 2 Q_q \kappa_{\text{NC}, q}\sw \right)
  \label{eq:gV-NLO} \,.
\end{align}

The CC cross section at LO is written as
\begin{equation}
  \frac{d^2\sigma^{\rm CC}(e^\pm p)}{dxd\Qsq}
  = \left(1 \pm P_e\right)
  \frac{\gf^2}{4\pi x}
  \left[\frac{m_W^2}{m_W^2+\Qsq}\right]^2
  \left(Y_+ W_2^\pm(x,\Qsq) \mp Y_{-} xW_3^\pm(x,\Qsq)
  - y^2 W_{\rm L}^\pm(x,\Qsq)\right)~. 
  \label{eq:cc-cs}
\end{equation}
In the quark-parton model, $W_{\rm L}^\pm = 0$, and the structure
functions $W_2^\pm$ and $xW_3^\pm$ are obtained from the parton
distribution functions. For electron scattering, only
positively charged quarks contribute:
\begin{equation}
  W_2^- =
  x \left( U + \overline{D} \right)
  \, ,
  \quad xW_3^- =
  x \left( U - \overline{D} \right)
  \, ,
  \label{eq:w23el-LO}
\end{equation}
while negatively charged quarks contribute to positron scattering:
\begin{equation}
  W_2^+ =
  x \left( \overline{U} + D \right)
  \, ,
  \quad xW_3^+ =
  x \left( D - \overline{U} \right)
  \, .
  \label{eq:w23po-LO}
\end{equation}
Below the top-quark threshold, one has
\begin{equation}
  U = u+c\, , \quad
  \overline{U} = \bar{u} + \bar{c}\, , \quad
  D = d+s\, , \quad
  \overline{D} = \bar{d} + \bar{s} \, .
\end{equation}

Higher-order EW corrections are collected in form factors
$\rho_{\text{CC}, eq/e\bar{q}}$. They modify the LO expressions
equations~\eqref{eq:w23el-LO} and \eqref{eq:w23po-LO}
as
\begin{align}
  W_2^- &=
  x \left( \rho^2_{\text{CC}, eq} U + \rho^2_{\text{CC},e\bar{q}} \overline{D} \right)
  \, ,
  \quad xW_3^- =
  x \left( \rho^2_{\text{CC},eq} U - \rho^2_{\text{CC},e\bar{q}} \overline{D} \right)
  \, ,
  \label{eq:w23el-NLO}
\\
  W_2^+ &=
  x \left( \rho^2_{\text{CC},eq} \overline{U}+ \rho^2_{\text{CC},e\bar{q}} D \right)
  \, ,
  \quad xW_3^+ =
  x \left( \rho^2_{\text{CC},e\bar{q}} D - \rho^2_{\text{CC},eq} \overline{U} \right)
  \, .
  \label{eq:w23po-NLO}
\end{align}

In the on-shell (OS) scheme~\cite{Sirlin:1980nh,Sirlin:1983ys},
the independent parameters of the SM EW theory are
determined by the fine structure constant $\alpha$ and the masses
of the gauge bosons, the Higgs boson $m_H$, and
the fermions $m_f$. The weak mixing angle is then fixed, and $\gf$ is a prediction, given by
\begin{equation}
  \gf=\frac{\pi\alpha}{\sqrt{2}m_W^2}
  \frac{1}{\sw}
  \frac{1}{(1-\dr)} \, , 
  \label{eq:gf-mw-mz}
\end{equation}
where higher-order corrections enter through the
quantity $\dr = \dr(\alpha, m_W, m_Z, m_H, m_t, \ldots)$
\cite{Sirlin:1980nh}, which describes corrections to the muon
decay beyond the tree-level~\cite{Bohm:1986rj,Hollik:1988ii}.

The $\rho_\text{NC}$, $\kappa_\text{NC}$ and $\rho_\text{CC}$ parameters are introduced to cover
the universal higher-order EW corrections described by
loop insertions in the boson propagators. The $\rho_\text{NC}$ parameters absorb
$Z$-boson propagator corrections combined with higher-order
corrections entering the $\gf$-$\mW$-$\sw$ relation, 
equation~\eqref{eq:gf-mw-mz}, while the $\kappa_\text{NC}$ parameters absorb 
one-loop $\gamma Z$ mixing propagator corrections. In addition,
there are higher-order corrections to the photon propagator
which can be taken into account by using the running fine structure
constant. Non-universal corrections due to vertex one-loop
Feynman graphs and box diagrams are added separately to the
NC cross sections. For the CC cross sections, both universal
and non-universal corrections can be combined into the form factors
$\rho_{\text{CC},eq/e\bar{q}}$.
The dominating corrections in this case are due
to loop insertions in the $W$-boson propagator.

\begin{figure}[tb]
  \begin{center}
    \includegraphics[width=0.495\textwidth]{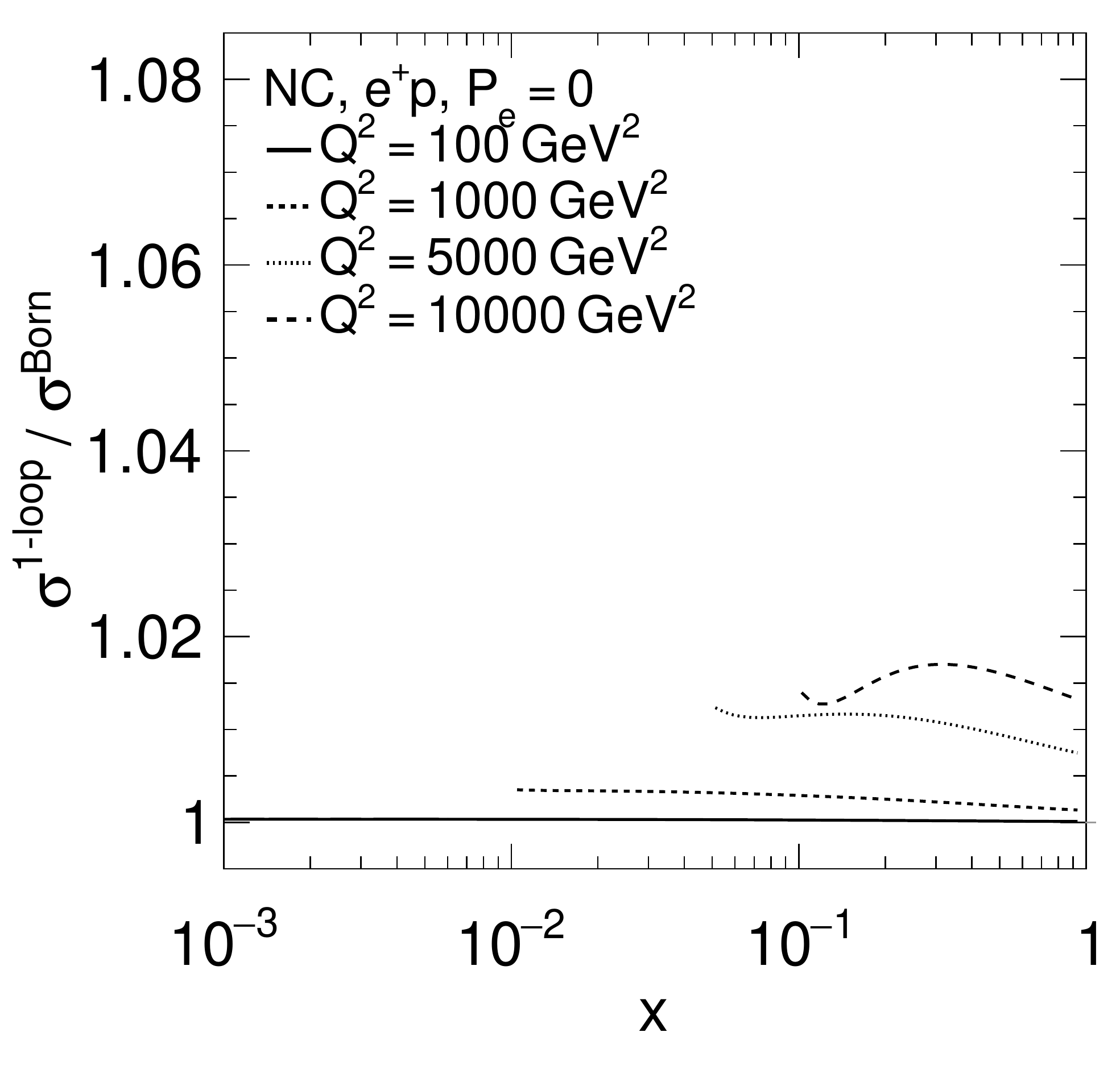%
}\hskip0.01\textwidth
    \includegraphics[width=0.495\textwidth]{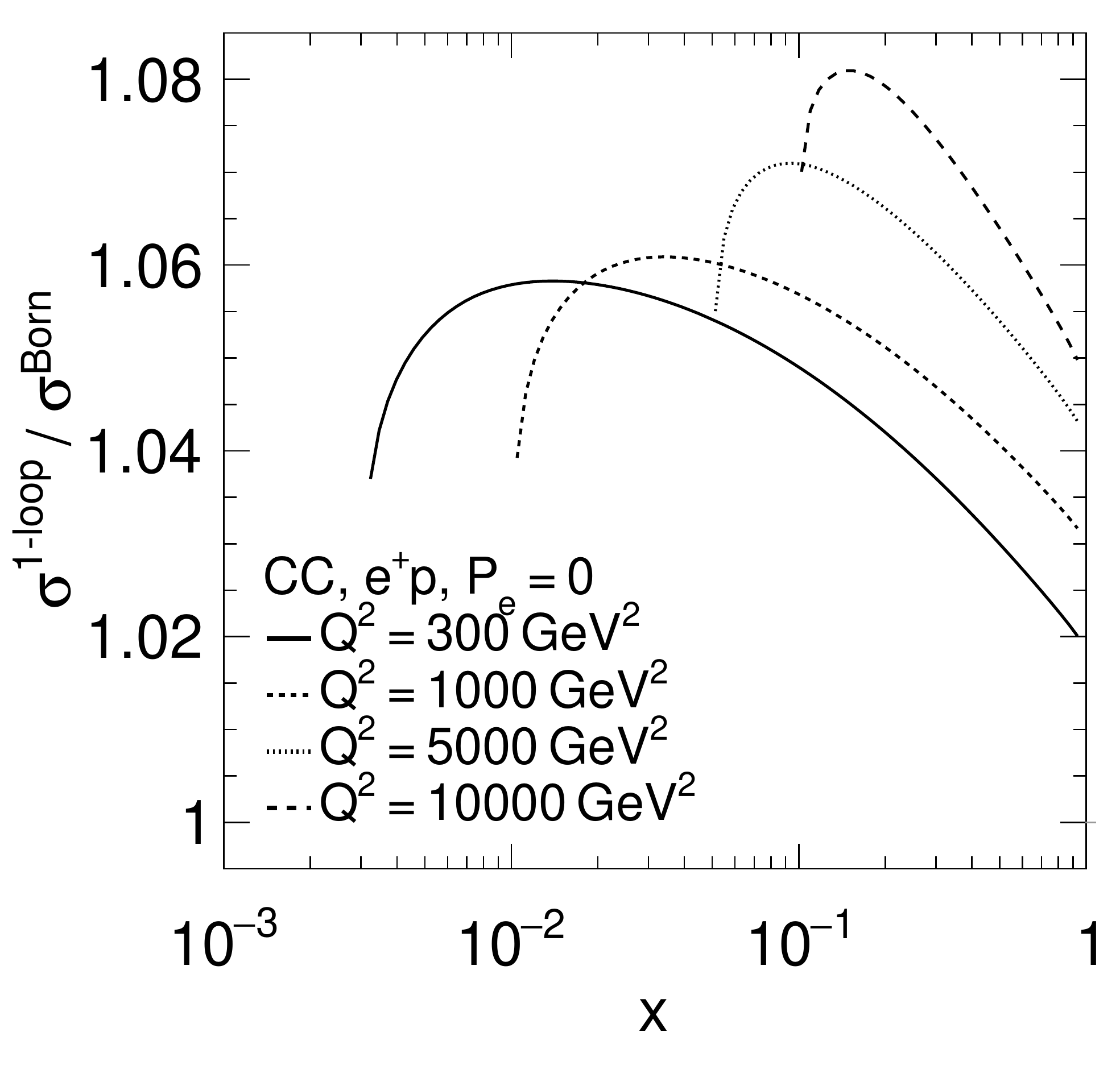}\hskip0.01\textwidth
  \end{center}
  \caption{
    Size of the purely weak one-loop corrections for the $e^+p$
    unpolarised inclusive NC DIS (left) and CC DIS (right) cross sections
    at selected values of \Qsq\ as a function of \xbj.
    QED corrections due to real and virtual photons and corrections from
    the vacuum polarisation (the running of $\alpha$) are not included.
    The corrections for electron scattering and for the
    case of non-vanishing lepton beam polarisation are all very  
    similar to the positron case, such that they differ by
    less than $0.01$ units.
  }
  \label{fig:EWcorrections}
\end{figure}

One-loop EW corrections have been calculated
in refs.~\cite{Bohm:1986na,Bardin:1988by,Hollik:1992bz}
for NC and in refs.~\cite{Bohm:1987cg,Bardin:1989vz} for CC
scattering (see also ref.~\cite{Heinemann:1998kk} for a study of
numerical results).
The present analysis uses the implementation
of EW higher-order corrections in the program EPRC described in ref.~\cite{Spiesberger:1995pr}.
The size of the purely weak one-loop
corrections to the
differential cross sections is displayed in
figure~\ref{fig:EWcorrections} for 
selected values of \Qsq\ for $e^+p$ scattering.
It includes the $\rho_\text{NC/CC}$ and
$\kappa_\text{NC}$ form factors, as well as contributions from vertex
and box graphs.
The corresponding higher order corrections for electron
scattering or for non-zero lepton beam polarisation
differ by less than $0.01$ units from the corrections shown in
figure~\ref{fig:EWcorrections}.  
Higher-order QED corrections due to real and virtual
emission of photons, as well as vacuum polarisation, i.e.\ the
running of the fine structure constant, also have to be taken
into account \cite{Kwiatkowski:1990es,Charchula:1994kf}.
These effects, however, had been considered for the cross section measurement  and are
therefore not included here.

In the OS scheme, used in this analysis,
the higher-order correction factors $\rho_\text{NC}$, $\kappa_\text{NC}$ and
$\rho_\text{CC}$ are calculated as a function of $\alpha$ and the
input mass values.
They depend quadratically on the top-quark
mass through $\Delta\rho_t\sim m_t^2$, and logarithmically on the
Higgs-boson mass, $\Delta\rho_H\sim\ln\left(m^2_H/\mW^2\right)$.
On the $Z$ pole they amount to about 4\%.
For DIS at HERA they are of similar size, but they exhibit a
non-negligible \Qsq-dependence~\cite{Spiesberger:1993jg}.
In  a modified version of the OS scheme~\cite{Marciano:1980pb},
commonly used in QCD analyses of DIS data, the Fermi constant can be
used to fix the input parameters replacing the $W$-boson mass as an
input parameter.
In that case the one-loop corrections are very small, 
i.e.\ $\rho_{\text{CC},eq/e\bar{q}}$ deviate from 1 by a few per mille.

Many extensions of the SM predict modifications of the weak 
NC couplings. They can be described conveniently 
by introducing additional parameters $\rhop{}$ and $\kapp{}$, thus
modifying the SM corrections.
Also for charged current cross sections, similar $\rhopW{}$ parameters
describing non-standard modifications of the CC couplings can be
introduced. 
The $\rhop{}$,  $\kapp{}$ and $\rhopW{}$ are introduced
through the following replacements in
equations~\eqref{eq:gA-NLO}, \eqref{eq:gV-NLO}, \eqref{eq:w23el-NLO} 
and \eqref{eq:w23po-NLO}:
\begin{align}
  \rho_\text{NC} &\rightarrow \rhop{}\rho_\text{NC}  \label{eq:rhozkapz1}~,\\
  \kappa_\text{NC} &\rightarrow \kapp{}\kappa_\text{NC}  \label{eq:rhozkapz2}~,\\
  \rho_\text{CC} &\rightarrow \rhopW{}\rho_\text{CC} \label{eq:rhopW}~.
\end{align}
In the SM, the parameters $\rhop{}$, $\kapp{}$ and $\rhopW{}$ are defined to be
1.
Various models with physics beyond the SM predict typical 
flavour-dependent deviations from 1 and therefore distinct parameters
for quarks ($\rhop{,q}$ and $\kapp{,q}$) and for leptons ($\rhop{,e}$ and $\kapp{,e}$) are considered.
These parameters may also depend on the energy scale. 
Precision EW measurements on the 
$Z$ resonance are sensitive to the NC couplings
at $m_Z$~\cite{ALEPH:2005ab}, while DIS is also probing their \Qsq\ dependence.
For CC there could be independent modifications (\rhopW{}) for
the lepton and quark couplings for each generation.
However, only the product of lepton times quark couplings appears in the final
expression for the cross section and therefore the same non-standard
coupling for all generations is assumed here.
Nonetheless, new 4-fermion operators can introduce
a difference between electron-quark and electron-antiquark scattering,
and thus two distinct parameters \rhopW{,eq}\ and \rhopW{,e\bar{q}}
are considered. 
These possibly scale-dependent parameters allow for additional 
tests of the SM couplings.

\section{H1 inclusive DIS cross section data}
\label{sec:data}
This study is based on the entire set of measurements of inclusive NC
and CC DIS cross sections by the H1 Collaboration, using data samples
for $e^+p$ and $e^-p$ taken in HERA-I and HERA-II.
The measurements are subdivided into two kinematic ranges,
corresponding to different subdetectors where the leptons with small
and large scattering angles are identified: low- and medium-\Qsq\ for
values of \Qsq\ typically 
smaller than $150\,\GeVsq$ and high-\Qsq for larger values up to 50\,000\,\GeVsq.
A summary of the data sets used is given in table~\ref{tab:table1}.
\begin{table}[tb]
  \footnotesize
  \centering
  \begin{tabular}{rl|cccccc}
   \hline
   \multicolumn{2}{c|}{Data set} & \Qsq-range & $\sqrt{s}$ & \Lumi & No. of & Polarisation & Ref.  \\
                             &  & [{\GeVsq}]   & [{\GeV}] & [{${\rm pb}^{-1}$}] &     data points   & [\%]  &   \\
   \hline
    1 & \ep\ combined low-\Qsq  & (0.5) 8.5 -- 150 & 301,319 & 20,\,22,\,97.6  &  94 (262) & --  &  \cite{Collaboration:2010ry} \\
    2 & \ep\ combined low-$E_p$ & (1.5) 8.5 -- 90  & 225,252 & 12.2,\,5.9 & 132 (136) & --  &  \cite{Collaboration:2010ry} \\

    3 & \ep\ NC 94--97  & 150 -- 30\,000 & 301 & 35.6 & 130 &  --& \cite{Adloff:1999ah} \\
    4 & \ep\ CC 94--97  & 300 -- 15\,000 & 301 & 35.6 &  25 &  -- & \cite{Adloff:1999ah} \\
   
    5 & \emm\ NC 98--99 & 150 -- 30\,000 & 319 & 16.4 & 126 & -- & \cite{Adloff:2000qj} \\
    6 & \emm\ CC 98--99 & 300 -- 15\,000 & 319 & 16.4 &  28 & -- & \cite{Adloff:2000qj} \\
   
    7 & \emm\ NC 98--99 high-$y$ & 100 -- 800 & 319 & 16.4 & 13 & -- & \cite{Adloff:2003uh} \\
    8 & \ep\ NC 99--00  & 150 -- 30\,000 & 319 & 65.2 & 147 & -- & \cite{Adloff:2003uh} \\
    9 & \ep\ CC 99--00  & 300 -- 15\,000 & 319 & 65.2 &  28 & -- & \cite{Adloff:2003uh} \\
   
    10 & \ep\  NC L HERA-II & 120 -- 30\,000 & 319 &  80.7  &  136 &  $-37.0 \pm 1.0$ &  \cite{Aaron:2012qi,Aaron:2012kn} \\
    11 & \ep\  CC L HERA-II & 300 -- 15\,000 & 319 &  80.7  &  28  &  $-37.0 \pm 1.0$ &  \cite{Aaron:2012qi,Aaron:2012kn} \\
    12 & \ep\  NC R HERA-II & 120 -- 30\,000 & 319 & 101.3  &  138 &  $+32.5 \pm 0.7$ &  \cite{Aaron:2012qi,Aaron:2012kn} \\
    13 & \ep\  CC R HERA-II & 300 -- 15\,000 & 319 & 101.3  &  29  &  $+32.5 \pm 0.7$ &  \cite{Aaron:2012qi,Aaron:2012kn} \\
    14 & \emm\ NC L HERA-II & 120 -- 50\,000 & 319 & 104.4  &  139 &  $-25.8 \pm 0.7$ &  \cite{Aaron:2012qi,Aaron:2012kn} \\
    15 & \emm\ CC L HERA-II & 300 -- 30\,000 & 319 & 104.4  &  29  &  $-25.8 \pm 0.7$ &  \cite{Aaron:2012qi,Aaron:2012kn} \\
    16 & \emm\ NC R HERA-II & 120 -- 30\,000 & 319 &  47.3  &  138 &  $+36.0 \pm 0.7$ &  \cite{Aaron:2012qi,Aaron:2012kn} \\
    17 & \emm\ CC R HERA-II & 300 -- 15\,000 & 319 &  47.3  &  28  &  $+36.0 \pm 0.7$ &  \cite{Aaron:2012qi,Aaron:2012kn} \\
    18 & \ep\  NC HERA-II high-$y$ &  60 -- 800 & 319 & 182.0  &   11  &  -- &  \cite{Aaron:2012qi,Aaron:2012kn} \\
    19 & \emm\ NC HERA-II high-$y$ &  60 -- 800 & 319 & 151.7  &   11  &  -- &  \cite{Aaron:2012qi,Aaron:2012kn} \\
   \hline
   \end{tabular}
  \caption{Data sets used in the combined EW and QCD fits. For
    each of the data sets, the corresponding range in \Qsq, the
    centre-of-mass energy $\sqrt{s}$, the corresponding integrated luminosity values, the number of measured data
    points, and the average longitudinal polarisation values of the lepton beam are given. 
    During the HERA-I running period data were taken with unpolarised lepton beams. 
    The numbers in brackets denote the respective quantities for the full data set, i.e.\ without the selection of $\Qsq\geq8.5\,\GeVsq$.
    The low- and medium-\Qsq\ data sets for $\sqrt{s} = 319$, 301, 252 and
    225\,GeV are combined into two common data sets as described in
    ref.~\cite{Collaboration:2010ry}.
    The data sets include electron and
    positron beams as well as neutral current (NC) and charged current
    (CC) cross sections.
    The data sets 10--17 are updated following the discussions in
    section~\ref{sec:data} and in appendix~\ref{appx:cs}.
  } 
  \label{tab:table1}
\end{table}

The low- and medium-\Qsq\ data sets (data sets 1 and 2)~\cite{Collaboration:2010ry} are
combined data sets, and they represent all corresponding NC DIS measurements
at different beam energies and during different data taking
periods published by
H1~\cite{Adloff:1997mf,Adloff:2000qk,Aaron:2009bp,Aaron:2009kv,Collaboration:2010ry}.
For these data photon exchange dominates over electroweak effects, but
they are important in this analysis to constrain the proton PDFs with
high precision.

Cross section measurements at high \Qsq\ are published separately for the
individual data taking periods (data sets: 3--4\,\cite{Adloff:1999ah}, 5--7\,\cite{Adloff:2000qj,Adloff:2003uh}, 8--9\,\cite{Adloff:2003uh}, 10--19\,\cite{Aaron:2012qi}).
The HERA-II data\footnote{The numerical values of the
  HERA-II cross sections~\cite{Aaron:2012qi} are corrected
  to the luminosity measurement erratum~\cite{Aaron:2012kn},
  by applying the factor 1.018.}
were taken with longitudinally polarised
lepton beams 
and exhibit smaller statistical uncertainties
due to the increased integrated luminosity, as compared to HERA-I.
The high-\Qsq\ data provide highest sensitivity for the
determination of the EW parameters.
The availability of longitudinally polarised lepton beams at HERA-II
further improves the sensitivity to the vector couplings $\vq$,  as
compared to unpolarised data.  
The data are restricted to $\Qsq\geq8.5\,\GeVsq$, for which 
quark mass effects are expected to be small, and
NNLO QCD predictions~\cite{Vogt:2004mw,Moch:2004pa} 
are expected to provide a good description of the
data~\cite{Abramowicz:2015mha,Andreev:2017vxu}.

All the data samples (data sets 1--19) had been corrected for higher-order QED 
effects due to the emission of photons from the lepton line, photonic
lepton vertex corrections, self-energy contributions 
at the external lepton lines, and fermionic contributions to the
running of the fine structure constant (c.f.\ ref.~\cite{Adloff:1999ah}). 
QED radiative corrections due to the exchange of two or more photons
between the lepton and the quark lines are small compared to the quoted
errors of the QED corrections and had been neglected (c.f.\ ref.~\cite{Adloff:2000qj}).
In the case of CC cross sections, the data had been corrected for
$\Ord(\alpha)$ QED effects at the lepton line (c.f.\ ref.~\cite{Adloff:1999ah}). 

In order to ensure that all first order EW corrections are considered
fully and consistently in this analysis, the applied QED corrections
to the input data are revisited in detail.
In the formulae for the cross section derivation~\cite{Aaron:2012qi},
the QED corrections are applied together with acceptance, resolution,
and bin-centre corrections, using two independent implementations of the cross section calculations.
It turns out that for the HERA-II data (data sets 10--19,
ref.~\cite{Aaron:2012qi}), these two implementations have employed
slightly different numerical values for the input EW parameters, and
furthermore have considered different components of the
higher-order EW corrections.
The corrections are therefore re-evaluated and updated values of
the previously published cross sections are obtained for this analysis.
The procedure is equivalent to the initial cross section
determination and therefore does not introduce additional uncertainties.
The updated cross sections for the data sets 10--17, as used in this
analysis, are provided in the appendix~\ref{appx:cs}.
The differences to the published cross sections are significantly
smaller than the statistical uncertainties for any data point.
The data sets 18 and 19 are at lower values of \Qsq\ and remain
unchanged, as well as the HERA-I data (data sets 1--9).
The effect of these updates is expected to be small for QCD 
analyses~\cite{Aaron:2012qi,Abramowicz:2015mha,Andreev:2017vxu}. 
As a cross check, fits similar to H1PDF2012 \cite{Aaron:2012qi} were performed
using either previously published data
\cite{Aaron:2012qi,Aaron:2012kn} or the corrected data given in the
appendix.
The two fits are in agreement within experimental
uncertainties, where the largest deviations of size one standard
deviation are observed for the down-valence contribution at low
factorisation scales.
In the present analysis the impact is also found to be insignificant,
but the updated cross sections are nevertheless applied in order to
have best consistency between data and the predictions used in the fits
described below.

\section{Methodology}
\label{section:methodology}
The EW parameters are determined in fits of the predictions
to data, where in addition to the EW parameters of interest also
parameters of the PDFs  are determined in order to account for PDF
uncertainties.
The fits are denoted according to their fit
parameters, for instance `\mW+PDF' denotes a determination of
\mW\ together with the parameters of the PDFs.

A dedicated determination of the PDFs in this analysis is important,
since all state-of-the-art PDF sets were determined using H1 data, while
assuming that the EW parameters take their SM values.
Hence, the use of such PDF sets could bias the results. 
Furthermore, PDF sets which include the H1 data suffer from the
additional complication that the same data were to be used twice, thus
leading to underestimated uncertainties.

The parameterisation of the PDFs 
follows closely the approach of
ref.~\cite{Abramowicz:2015mha}, where the PDF set
HERAPDF2.0\footnote{
HERAPDF2.0 is determined from combined inclusive NC and CC data from
the H1 and ZEUS experiments assuming unpolarised lepton beams.
} 
was obtained, using EW parameters determined from other experiments. 
The parameterisation uses five functional forms with altogether 13 fit
parameters, defined at the starting scale $Q_0^2=1.9\,\GeVsq$.
The scale dependence of the PDFs is evaluated using the DGLAP formalism.

As opposed to the HERAPDF2.0 analysis, 
the {\sc Alpos} fitting framework~\cite{Andreev:2017vxu} is used in
the present analysis.
The cross section predictions have been validated against the xFitter
framework~\cite{Aaron:2012qi,Alekhin:2014irh,Abramowicz:2015mha},
which is the successor of the H1Fitter framework~\cite{Aaron:2009kv}. 
The structure functions are obtained in the zero-mass
variable-flavour-number-scheme at NNLO in QCD using the 
QCDNUM code~\cite{Botje:2010ay,Botje:2016wbq}.
The one-loop EW corrections are included in an updated version of the
EPRC code~\cite{Spiesberger:1995pr}, while the data have already
been corrected for higher-order QED radiative 
effects, as outlined in section~\ref{sec:data}.

The goodness of fit,
\chisq, is derived from a likelihood function assuming the quantitites
to be normal distributed in terms of relative uncertainties~\cite{Andreev:2014wwa,Andreev:2017vxu},
which is equivalent to log-normal distributed quantities in terms of
absolute uncertainties.
The log-normal distribution is strictly positive and a good
approximation of a Poisson distribution.
The latter is important, since in the kinematic domain where the data
exhibit the highest sensitivity to the EW parameters, the statistical
uncertainties may become sizeable and dominating.  
The \chisq\ is calculated as
\begin{equation}
  \chi^2 = \sum_{ij}\log\tfrac{\varsigma_i}{\tilde\sigma_i} V_{ij}^{-1}\log\tfrac{\varsigma_j}{\tilde\sigma_j}\,,
\end{equation}
where the sum runs over all data points with measured cross sections 
$\varsigma_{i}$ and the corresponding theory predictions,
$\tilde\sigma_i$. The covariance matrix $V_{ij}$ is constructed from all
relative uncertainties, taking also correlated uncertainties between 
the data sets into account~\cite{Aaron:2012qi}.
The beam polarisation measurements provide four additional 
data points, included in the vector $\varsigma$, with their
uncertainties~\cite{Sobloher:2012rc} and four corresponding 
parameters in the fit.

The PDF fit alone, i.e.\ all EW parameters set to their SM 
values~\cite{Olive:2016xmw}, 
yields a fit quality of
$\chisq/\ndf = 1432 / (1414-17)=1.03$, 
where the number of degrees of freedom, $n_\text{dof}$, is calculated
from 1410 cross section data points plus 4 measurements of the polarisation, 
and considering 13 PDF and 4 fit polarisation parameters.
This indicates an overall good description of the data by the employed
model.
More detailed studies of the QCD analysis with the given data
samples have been presented previously~\cite{Aaron:2012qi,Andreev:2017vxu}.

\section{Results} 
This section reports the results of different fits, starting with mass
determinations in section~\ref{sec:mass}, followed by weak NC coupling
determinations in section~\ref{sec:couplings} and the study of
$\rhop{}$, $\kapp{}$ and $\rhopW{}$ parameters in section~\ref{sec:ff}.

\subsection{Mass determinations}\label{sec:mass}
The masses of the $W$ and $Z$ bosons, as well as the top-quark mass
are determined using different prescriptions to fix the fit parameters of the EW theory in the OS 
scheme. The different prescriptions lead to different sensitivities of the measured cross sections
to the EW parameters~\cite{Blumlein:1987fd}.
The results are summarised in table~\ref{tab:masses}.
\begin{table}[hbt]
  \footnotesize
  \centering
   \begin{tabular}{lc@{$\,=\,$}lc}
   \hline
   Fit parameters & \multicolumn{2}{c}{Result} & Independent input parameters \\
   \hline 
   \mW+PDF  & $\mW$ & $80.520\pm0.070_{\rm stat}\pm 0.055_{\rm syst}\pm0.073_{\PDF}\,\GeV$ & $\alpha$, $m_Z$, $m_t$, $m_H$, $m_f$ \\ 
   \mWprop+PDF & $\mWprop $ & $ 80.62\pm0.67_{\rm stat}\pm0.17_{\rm syst}\pm0.38_{\PDF}\,\GeV$ & $\alpha$, $m_W$, $m_Z$, $m_t$, $m_H$, $m_f$ \\
   \mWGfW+PDF &  $\mWGfW $ & $ 82.05\pm0.51_{\rm stat}\pm0.44_{\rm syst}\pm0.37_{\PDF}\,\GeV$ & $\alpha$, \gf, $m_t$ $m_H$, $m_f$ \\
   \mZ+PDF & $\mZ $ & $ 91.084\pm0.064_{\rm stat}\pm 0.050_{\rm syst}\pm0.070_{\PDF}\,\GeV$ & $\alpha$, $m_W$, $m_t$, $m_H$, $m_f$ \\ 
   \mt+PDF & $\mt $ & $ 154\pm10_{\rm stat}\pm12_{\rm syst}\pm15_{\rm PDF}\pm15_{\mW}\,\GeV$ & $\alpha$, $m_W$, $m_Z$, $m_H$, $m_f$ \\ 
   \hline 
   \end{tabular}
   \caption{Results for five combined fits of mass parameters together with PDFs. 
   The multiple uncertainties correspond to statistical (stat), experimental 
   systematic (syst) and PDF uncertainties. The  $m_t$ determination
   also includes an uncertainty due to the uncertainty of the $W$
   mass. The most-right column lists further input parameters not
   varied in the fit.}
   \label{tab:masses}
\end{table}

\begin{figure}[tb]
  \begin{center}
    \includegraphics[width=0.48\textwidth]{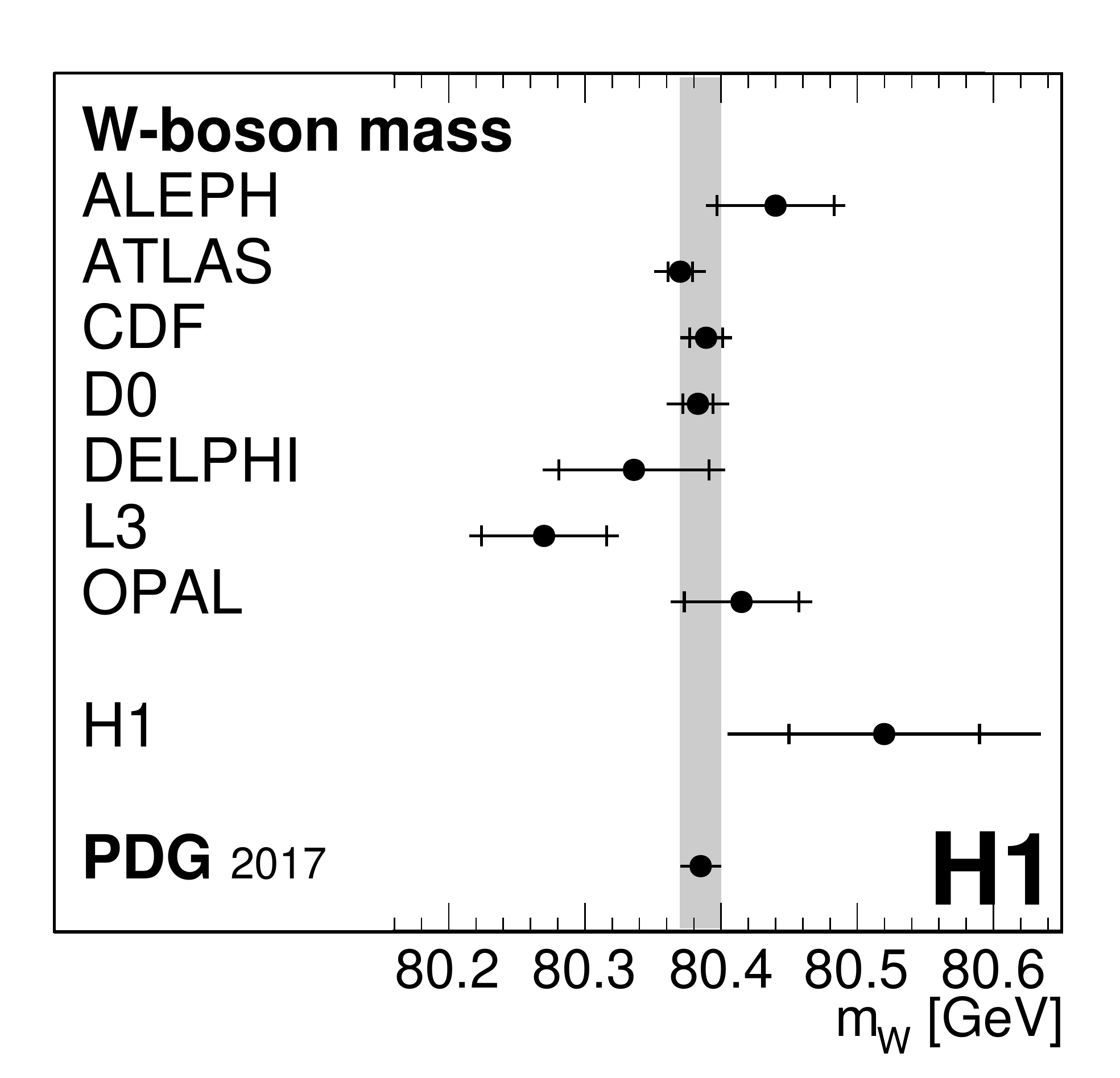
}\hskip0.01\textwidth
  \end{center}
  \caption{
    Value of the $W$-boson mass compared to results obtained by the ATLAS, ALEPH, CDF,
    D0, DELPHI, L3 and OPAL experiments, and the world average value.
    The inner error bars indicate statistical uncertainties and the
    outer error bars full uncertainties.
}
\label{fig:mW}
\end{figure}

In the combined \mW+PDF fit, 
where 
$\alpha$, \mZ, \mt, \mH\ and $m_f$ are taken as external input values~\cite{Olive:2016xmw}, the EW parameter \mW\ is determined to be
\begin{equation}
  \mW = 80.520\pm0.070_{\rm stat}\pm0.055_{\rm syst}\pm0.074_{\PDF}=80.520\pm 0.115_{\rm tot}\,\GeV~.
\end{equation}
and the expected uncertainty\footnote{The expected
  uncertainty is obtained from a re-fit using the Asimov data
  set and the data uncertainties~\cite{Cowan:2010js}.}
is 0.118\,\GeV.
The total (tot) uncertainty is improved by
about a factor of two in comparison to the earlier result based on
HERA-I data only~\cite{Aktas:2005iv}.
The uncertainty decomposition is derived by switching off the
uncertainty sources subsequently or repeating the fit with fixed PDF parameters\footnote{ 
    The PDF uncertainty contains both a statistical and a 
    systematic component, but the systematic component 
    dominates.}.
Other uncertainties due to the input masses
(\mZ, \mt, \mH) and theoretical uncertainties, e.g.\ from 
incompletely known higher-order terms in \dr,
or model and parameterisation uncertainties of the
PDF fit, are all found to be negligible with respect to the
experimental uncertainty.
The correlation of \mW\ with any of the PDF parameters is weak,
with absolute values of the correlation coefficients below 0.2.
The global correlation coefficient~\cite{James:1975dr} of \mW\ 
in the EW+PDF analysis is 0.64.
The \mW\ sensitivity arises predominantly from the CC data, with the most
important constraint being the normalisation through \gf\ (see
equations~\eqref{eq:cc-cs} and \eqref{eq:gf-mw-mz}).
The highest sensitivity of the H1 data to \mW\ is at a \Qsq\ value of
about 3800\,\GeVsq.
The result for \mW\ is compared to determinations from other single
experiments~\cite{Barate:1997bf,Acciarri:1997ra,Ackerstaff:1996nk,Abdallah:2008ad,Abazov:2012bv,Aaltonen:2012bp,Aaltonen:2013vwa,Aaboud:2017svj}
in figure~\ref{fig:mW}, and is found to be consistent with these
as well as with the world average value
of $80.385\pm0.015\,\GeV$~\cite{Aaltonen:2013iut,Olive:2016xmw}.
The $W$-mass determination in the space-like regime at HERA can be interpreted
as an indirect constraint on
\gf\ through equation~\eqref{eq:gf-mw-mz}, however in a process at large momentum
transfer.
Using the world average value of
\mZ~\cite{ALEPH:2005ab,Olive:2016xmw},
the result obtained here, $\mW=80.520\pm 0.115\,\GeV$,
represents an indirect determination of the weak mixing angle in the
OS scheme as
$\sw=0.22029\pm0.00223$. 
The uncertainty of the present $m_W$ determination matches the
anticipated HERA results in \cite{Blumlein:1987fd} and in
\cite{Beyer:1995pw,CooperSarkar:1998ug}.

Alternative determinations of \mW\ are also explored.
One option is to use exclusively
the dependence of the CC cross section on the propagator mass
$\sigma^{\rm CC}\propto\left(\mW^2/(\mW^2+\Qsq)\right)^2$.  
The result is $\mWprop=80.62\pm0.79\,\GeV$, with an expected
uncertainty of 0.80\,\GeV.
This improves the precision of the corresponding fit to HERA-I
data~\cite{Aktas:2005iv} by more than a factor of two. 
The value is consistent with the world average value and with the
result of the \mW+PDF fit.

Another \mW\ determination is based on
the high precision measurement of \gf~\cite{Tishchenko:2012ie}, which is
performed at low energy, together
with $\alpha$ as main external input.  
For this fit, \mZ\ is a prediction and is given by the \gf-\mW-\mZ\ relation in
equation~\eqref{eq:gf-mw-mz}.
With the precise knowledge of \gf, the normalisations of the CC predictions are known, and
therefore the predominant sensitivity to \mW\ arises from the $W$-boson
propagator, and the \mW\ dependence through \mZ\ in the NC
normalisation is small.
In this fit, the value of \mW, denoted as \mWGfW, is determined as
$\mWGfW=82.05\pm0.77\,\GeV$.
The value is consistent at about 2 standard deviations with the world
average value and with the result of the \mW+PDF fit above.
The larger uncertainty compared to the fit described above is expected.
This indirect determination of the $W$-boson mass assumes the validity of the SM~\cite{CooperSarkar:1998ug}.

\begin{figure}[tbh]
  \begin{center}
    \includegraphics[width=0.48\textwidth]{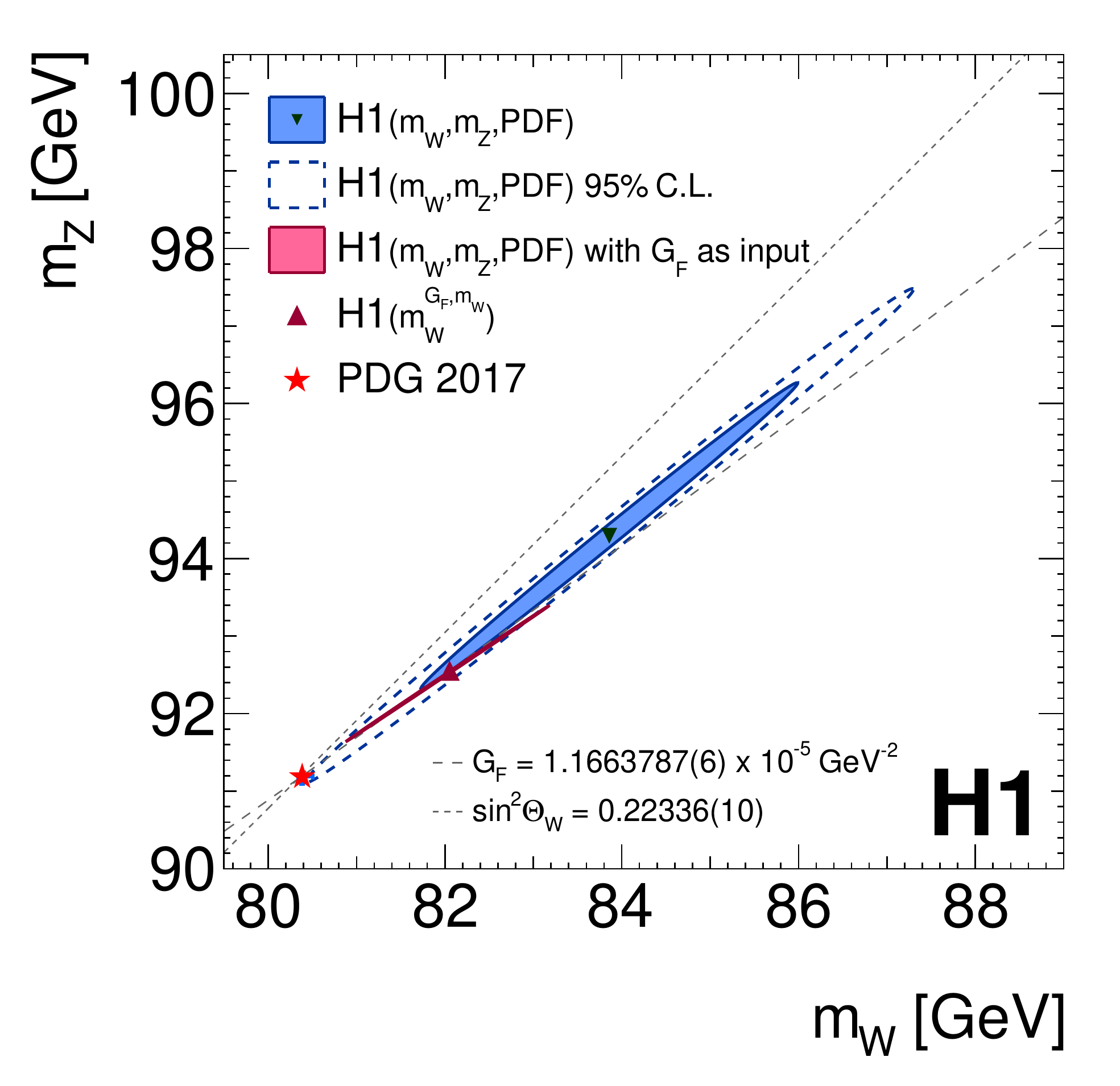%
}\hskip0.01\textwidth
    \includegraphics[width=0.48\textwidth]{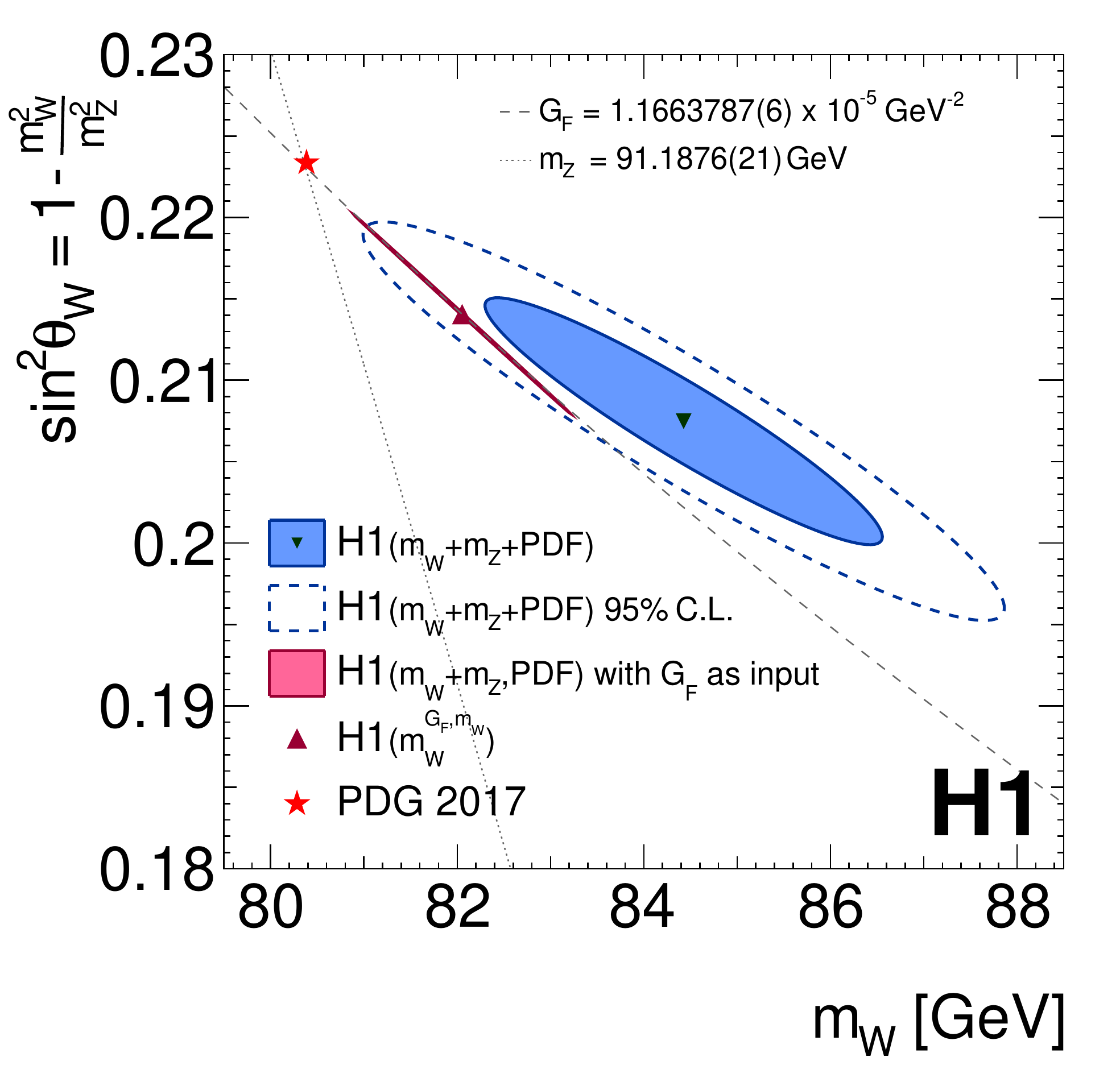}\hskip0.01\textwidth
  \end{center}
  \caption{
    Results of the \mW+\mZ+PDF fit, and the \mW+\mZ+PDF fit
    with \gf\ as additional input.
    For better visibility, the right panel displays the quantity
    $\sw=1-\mW^{2}/\mZ^{2}$ on the vertical axis and identical results as the left panel.
    The 68\,\% confidence level (C.L.) contour of the fit including the
    \gf\ measurement is very shallow.
    The result of the \mWGfW\ fit is further indicated but without uncertainties.
  }
\label{fig:mWmZ}
\end{figure}

A simultaneous determination of \mW\ and \mZ{} is also performed. 
The 68\,\% and 95\,\% confidence level contours of that \mW+\mZ+PDF fit
are displayed in figure~\ref{fig:mWmZ} (left).
Sizeable uncertainties $\Delta\mW=1.4\,\GeV$ and $\Delta\mZ=1.3\,\GeV$ with
a very strong correlation are observed.
A less strong correlation is found when displaying $\sw=1-\mW^2/\mZ^2$
instead  of \mZ\ (figure~\ref{fig:mWmZ}, right).
A mild tension of less than 3 standard deviations between the
world average values for \mW\ and \mZ\ and the fit result is observed.
The very strong correlation prevents a meaningful simultaneous
determination of the two boson masses from the H1 data alone.

In such a simultaneous determination of two mass parameters, the precise measurement
of \gf\ can be taken as additional input.
Due to its great precision it effectively behaves like a constraint,
as was proposed earlier~\cite{Brisson:1991vj,Spiesberger:1993jg}.
The 68\% confidence level contours of the \mW+\mZ+PDF fit with
\gf\ as one additional input data~\cite{Tishchenko:2012ie}, is further displayed in
figure~\ref{fig:mWmZ}.
As expected, the resulting value of \mW\ is equivalent to the value
obtained in the \mWGfW+PDF fit.
The 68\% confidence level contour is very shallow due to the high
precision of \gf.
The mild tension with the world average values of \mW\ and \mZ\ is 
reduced in comparison to the fit without \gf\ constraint.
In the \mW-\mZ\ plane the \gf\ constraint corresponds to a thin band.
The orientation of the \mW+\mZ+PDF contour is similar to the slope of
the \gf\ band, because the predominant sensitivity to \mW\ and \mZ\ of the H1
data arises through terms proportional to \gf\ and \sw\, rather than
the propagator terms.
This explains the large uncertainty observed in the \mWGfW+PDF fit as
compared to the nominal \mW+PDF fit.

The value of \mZ\ is determined in the \mZ+PDF fit to 
$\mZ=91.08\pm0.11\,\GeV$, to be compared with the
measurements at the $Z$ pole of $\mZ=91.1876\pm0.0021\,\GeV$~\cite{ALEPH:2005ab}.
The precision is very similar to the $W$-mass determination, as can
be expected from figure~\ref{fig:mWmZ}.

The value of \mt\ is determined in the \mt+PDF fit, where \mW\ and 
\mZ\ are taken as external input, yielding
 $ \mt=154\pm10_{\rm stat}\pm12_{\rm syst}\pm15_{\rm PDF}\pm15_{\mW}\,\GeV$.
The last uncertainty accounts for the
$W$-mass uncertainty of $15\,\MeV$~\cite{Olive:2016xmw}.
The result is consistent with direct measurements at the
LHC~\cite{Aad:2015nba,Khachatryan:2015hba,Sirunyan:2017idq,Aaboud:2017mae,CMS:2017eoz}
and Tevatron~\cite{TevatronElectroweakWorkingGroup:2016lid}.
At HERA, the top quark mass contributes only through loop effects,
this explains the moderate sensitivity and the strong dependence on
the $W$ mass.

Higher-order corrections to \gf\ (see equation~\eqref{eq:gf-mw-mz}, \dr)
include bosonic self-energy corrections~\cite{Marciano:1980pb}
with a logarithmic dependence on the Higgs-boson mass, \mH, and thus
could, in principle, allow for constraints on \mH~\cite{Blumlein:1987fd}.
At HERA, however, the Higgs-boson mass dependent contribution is too small and 
no meaningful constraints on \mH\ can be obtained with the HERA data.  

A further study on the determination of EW parameters is
performed, by considering the precision measurements of
\mZ~\cite{ALEPH:2005ab}, \gf~\cite{Tishchenko:2012ie},
\mt~\cite{Olive:2016xmw} and \mH~\cite{Aad:2015zhl} as
experimental input data in addition to the H1 data.
In this simplified global fit, it is observed that the H1 data cannot
provide 
significant constraints, 
for instance on the $W$-boson mass or its correlation to any other
parameter.
This is because a precision of 7\,MeV on \mW\ is already achieved
through indirect
constraints~\cite{Olive:2016xmw,deBlas:2016ojx,Haller:2018nnx}.

\subsection{Weak neutral-current couplings }\label{sec:couplings}

\begin{figure}[tb]
  \begin{center}
    \includegraphics[width=0.48\textwidth]{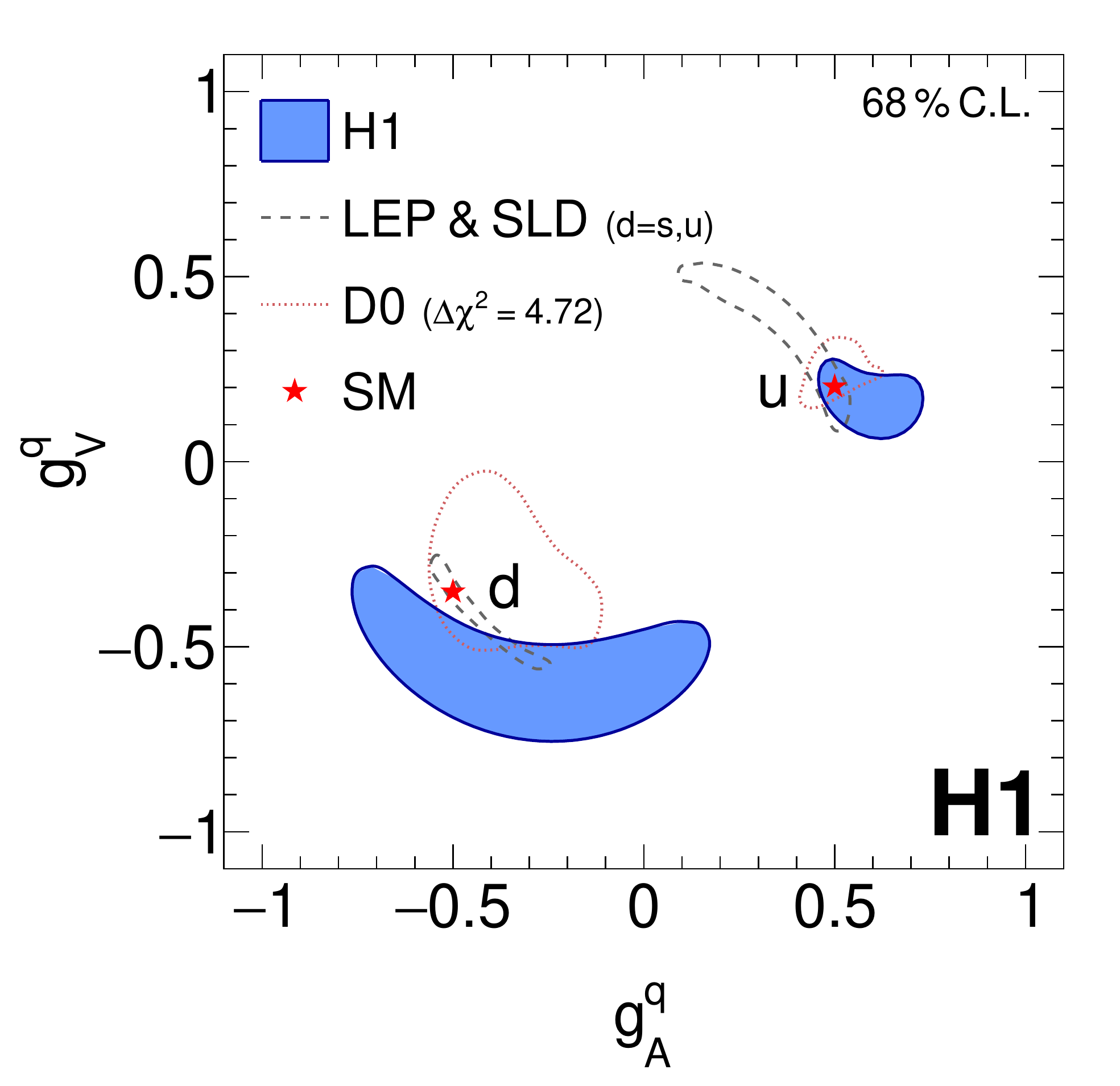%
}\hskip0.01\textwidth
    \includegraphics[width=0.48\textwidth]{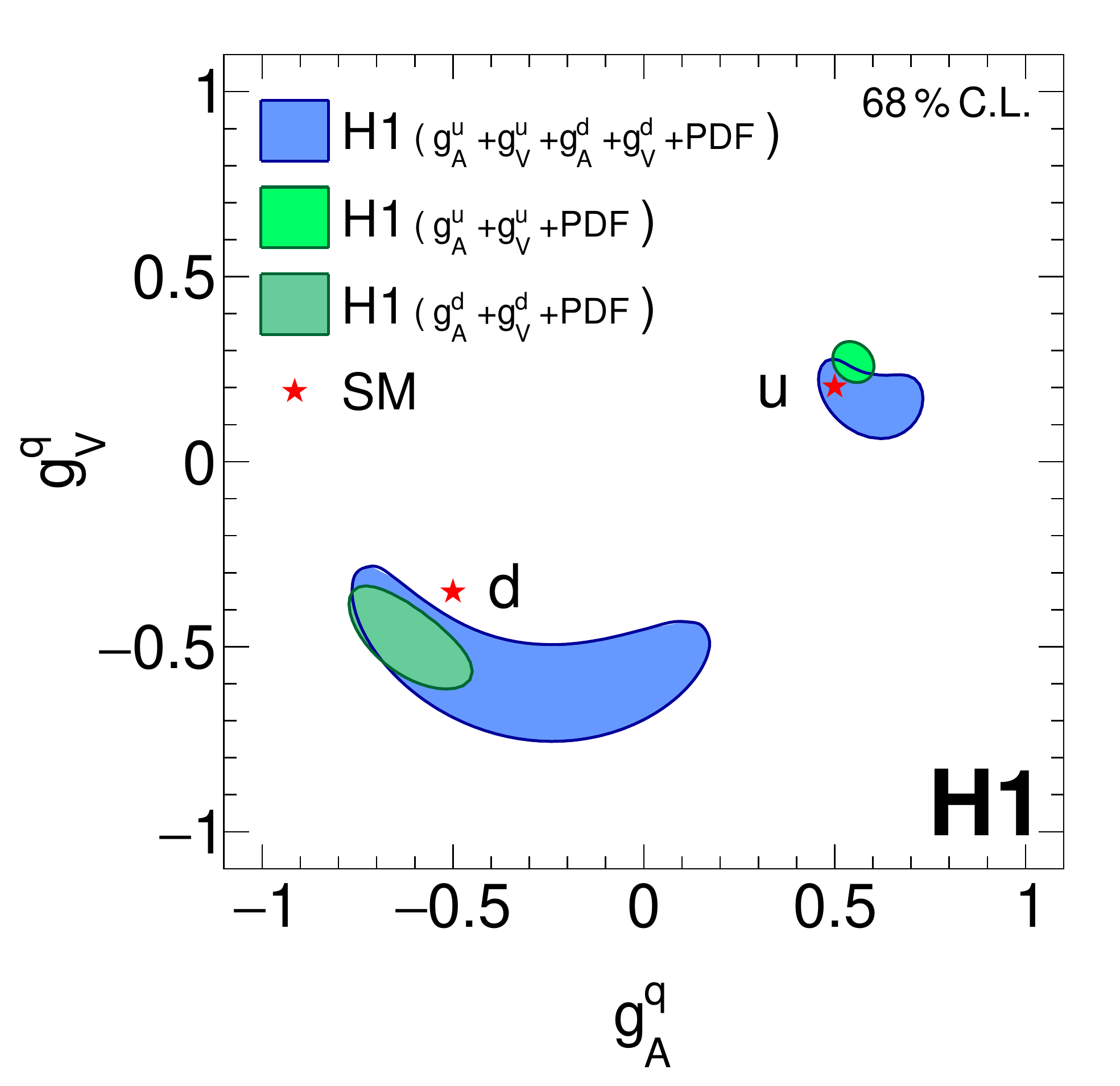}\hskip0.01\textwidth
  \end{center}
  \caption{
  Results for the weak neutral-current couplings of the $u$- and $d$-type
  quarks at the 68\% confidence level~(C.L.) obtained with the
  \au+\vu+\ad+\vd+PDF fit.
  The left panel shows a comparison with results from the D0, 
  LEP and SLD experiments (the mirror solutions are not
  shown). 
  The 68\%~C.L.\ contours of the H1 results correspond to
  $\Delta\chi^2 = 2.3$,  
  where at the contour all other fit parameters are minimised. 
  The SM expectation is displayed as a star.
  The right panel shows a comparison of results from fits 
  where the couplings of one quark type are fit parameters, and 
  the couplings of the other quark type are fixed, i.e.\ the
  \au+\vu+PDF and \ad+\vd+PDF fits. 
}
\label{fig:couplings}
\end{figure}

The weak NC couplings, defined in
equations~\eqref{eq:gA-NLO} and \eqref{eq:gV-NLO}, enter the
calculation of the structure functions in equations~\eqref{eq:last1}
and \eqref{eq:last2}. 
They are scale dependent beyond the tree-level approximation.
The fit parameters for the axial-vector and vector couplings considered 
here are defined as the tree-level parameters, given in 
equations~\eqref{eq:gA-LO} and \eqref{eq:gV-LO}.
The one-loop corrections are taken into account 
through multiplicative factors.
Results of the fits thus are 
compared with the SM tree-level predictions for the axial-vector and 
vector coupling constants. 
The axial-vector and vector couplings of the $u$- and
$d$-type quarks, $g_A^{u/d}$ and $g_V^{u/d}$, are determined in a 
combined fit together with the PDF parameters
and the results are presented in table~\ref{tab:couplings}. 
The two-dimensional contours representing the 68\% confidence level for two fit parameters are displayed and 
compared\footnote{It is worth to note that the results
  are corrected to the Born-level, whereas other experiments 
  often consider effective couplings defined at the 
  $Z$ pole~\cite{ALEPH:2005ab,Abazov:2011ws}. Such a fixed-scale 
  definition of couplings is not suitable for DIS, where data 
  cover a wide range of \Qsq values. On the other hand, the relation 
  between tree-level and effective $Z$-pole couplings is well known 
  (see for example \cite{ALEPH:2005ab}), and the differences of 
  corresponding numerical values are significantly
  smaller than the achieved precision.} 
with results from other experiments in figure~\ref{fig:couplings} (left).
The results are consistent with the SM expectation.
The sensitivity on \au\ and \vu\ is similar to LEP and D0 measurements.
The HERA measurements do not exhibit sign ambiguities or
ambiguities between axial-vector and vector couplings, which are for
example present in determinations from $Z$-decays at the pole.

The results for $g_A^{u/d}$ and $g_V^{u/d}$ obtained from this
analysis are found to be compatible
with fits, where alternatively 
external PDFs, such as ABMP16~\cite{Alekhin:2017kpj},
CT14~\cite{Dulat:2015mca}, H1PDF2017~\cite{Andreev:2017vxu}, MMHT14~\cite{Harland-Lang:2014zoa} or
NNPDF3.0~\cite{Ball:2014uwa}, are used and the corresponding PDF uncertainties
are considered in the \chisq\ definition.
As explained in Section \ref{section:methodology}, this approach yields underestimated uncertainties, but provides a valuable cross check.

\begin{table}[tb]
  \footnotesize
  \centering
   \begin{tabular}{lr@{\,=\,}r@{\,$\pm$\,}crrrr}
   \hline
   Fit parameters & 
   \multicolumn{3}{c}{Result} & \multicolumn{4}{c}{Correlations} \\   
   \multicolumn{4}{c}{} & \multicolumn{1}{c}{\au} & \multicolumn{1}{c}{\vu} & \multicolumn{1}{c}{\ad} & \multicolumn{1}{c}{\vd} \\
   \hline
   \au+\vu+\ad+\vd+PDF
     & \au &$ 0.614$ & 0.100  & 1.00    \\
     & \vu &$ 0.145$ & 0.056  &$-0.10$&  1.00   \\
     & \ad &$-0.230$ & 0.350  & 0.94  &$-0.10$&  1.00 &  \\ 
     & \vd &$-0.643$ & 0.083  &$ 0.13$&  0.70 &$-0.09$&  1.00   \\
   \hline 
   \au+\vu+PDF
     & \au & 0.548 & 0.036  & 1.00    &      &    &   \\
     & \vu & 0.270 & 0.037  & $-0.18$ & 1.00 &    &   \\
   \hline
   \ad+\vd+PDF
     & \ad & $-0.619$ & 0.108  &   &   &   1.00   &    \\
     & \vd & $-0.488$ & 0.092  &   &   &  $-0.68$ & 1.00  \\
   \hline
   \end{tabular}
     \caption{Results of the fitted weak neutral-current couplings of 
     the $u$- and $d$-type quarks. The other parameters $\alpha$, $m_W$, 
     $m_Z$, $m_t$, $m_H$ and $m_f$ are taken as external input~\cite{Olive:2016xmw}.
     The uncertainties quoted correspond to the total uncertainties.}
  \label{tab:couplings}
\end{table}

By extracting the couplings of the $u$- and $d$-type quarks
separately, i.e.\ fixing the couplings of the other quark type to
their SM expectations and performing a \au+\vu+PDF or \ad+\vd+PDF fit,
the uncertainties reduce significantly due to
weaker correlations between the fitted quark couplings.
The 68\% confidence level contours are also displayed in
figure~\ref{fig:couplings} (right), and numerical values are listed in table~\ref{tab:couplings}.

\begin{figure}[tb]
  \begin{center}
    \includegraphics[width=0.48\textwidth]{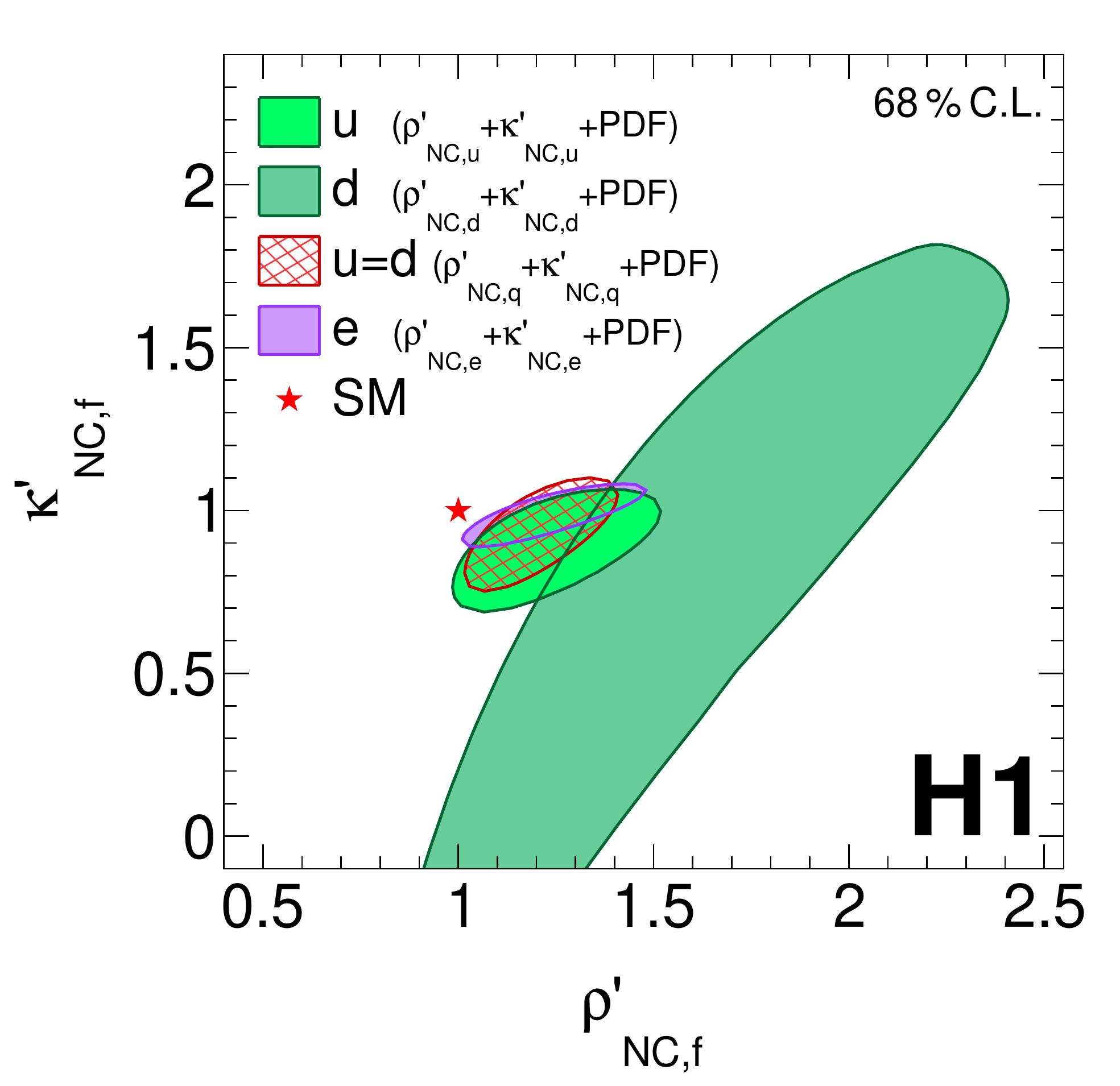
}\hskip0.01\textwidth
  \end{center}
  \caption{
    Results for the $\rhop{,f}$ and $\kapp{,f}$ parameters for $u$-
    and $d$-type quarks and electrons
    at 68\% confidence level (C.L.), obtained with the
    \rhop{,u}+\kapp{,u}+PDF, 
    \rhop{,d}+\kapp{,d}+PDF and
    \rhop{,e}+\kapp{,e}+PDF fits, respectively.
    The SM expectation is displayed as a star.
    The contour of the $d$-type quark is truncated due to the limited
    scale of the panel.
    For comparison, also the result of the \rhop{,q}+\kapp{,q}+PDF
    fit is displayed, where quark universality is assumed ($u=d$).
    The results of the \rhop{,u}+\kapp{,u}+PDF and
    \rhop{,d}+\kapp{,d}+PDF fits are equivalent to
    the \au+\vu+PDF and \ad+\vd+PDF fits, respectively, displayed
    in figure~\ref{fig:couplings}.
}
\label{fig:rhopkapp}
\end{figure}

\begin{boldmath}
\subsection{The $\rhop{}$, $\kapp{}$ and $\rhopW{}$ parameters}\label{sec:ff}
\end{boldmath}
The values of the $\rhop{,f}$ and $\kapp{,f}$ parameters
(c.f.\ equations~\eqref{eq:rhozkapz1} and~\eqref{eq:rhozkapz2}) are determined
for $u$- and $d$-type quarks and for electrons in 
\rhopu+\kappu+PDF, \rhopd+\kappd+PDF and \rhope+\kappe+PDF
fits, respectively.
In these fits, the respective \rhop\ and \kapp\ parameters are free
fit parameters, while the other $\rho^{\prime}$ and \kapp\ parameters
are set to one and the SM EW parameters are fixed.
Scale-dependent quantities such as $\rho_{NC,f}$, $\kappa_{NC,f}$,
$\rho_{CC,f}$ are calculated in the OS scheme as outlined in
section \ref{text:section:theory}.
The results are presented in table~\ref{tab:rhopzkap} and the 68\%
confidence level contours for the individual light quarks and for
electrons are shown in figure~\ref{fig:rhopkapp}.
The results are compatible with the SM expectation at 1--2 
standard deviations. 
The parameters of the $d$-type quarks exhibit larger uncertainties
than those of the $u$-type quarks.
This 
is due to the small electric charge
of the $d$ quark in the leading  $\gamma Z$-interference term (see
equations~\eqref{eq:last1} and~\eqref{eq:last2}),
and also in \vd\ (see equation~\eqref{eq:gV-NLO}).
Furthermore, the $d$-valence component of the PDF is smaller than the
$u$-valence component.

\begin{table}[tb]
  \footnotesize
  \centering
   \begin{tabular}{lccrc}
   \hline
   Fit parameters & \multicolumn{2}{c}{Result} & Correlation \\
   \hline
   \rhopu+\kappu+PDF & $\rhopu=1.23\pm0.17$ & $\kappu=0.88\pm0.12$ & 0.61 \\
   \rhopd+\kappd+PDF & $\rhopd=1.54\pm0.55$ & $\kappd=0.74\pm0.85$ & 0.92\\
   \rhop{,e}+\kapp{,e}+PDF & $\rhop{,e}=1.22\pm0.13$ & $\kapp{,e}=0.98\pm0.06$ & 0.74\\
   \hline
   \rhopd+\kappd+\rhopu+\kappu+PDF
      & \multicolumn{2}{c}{see appendix~\ref{appx:results}}  & \\
   \hline
   \rhop{,q}$+$\kapp{,q}+PDF & $\rhop{,q}=1.20\pm0.13$ & $\kapp{,q}=0.93\pm0.11$ & 0.69\\
   \hline
   \rhop{,q}+\kapp{,q}+\rhop{,e}+\kapp{,e}+PDF
      & \multicolumn{2}{c}{see appendix~\ref{appx:results}}  & \\
   \hline
   \rhop{,f}$+$\kapp{,f}+PDF & $\rhop{,f}=1.09\pm0.07$ & $\kapp{,f}=0.98\pm0.05$ & 0.83\\
   \hline
   \end{tabular}
   \caption{
     Results for $\rhop{}$ and $\kapp{}$ parameters and their
       correlation coefficients.
       The parameters $\alpha$, $m_W$, 
       $m_Z$, $m_t$, $m_H$ and $m_f$ are set to their SM values.
       The uncertainties quoted correspond to the total uncertainties.
   }
     \label{tab:rhopzkap}
\end{table}

The results of the \rhop{,u}+\kapp{,u}+PDF and
\rhop{,d}+\kapp{,d}+PDF fits (table~\ref{tab:rhopzkap}) are equivalent to
the values determined for the NC couplings in \au+\vu+PDF and
\ad+\vd+PDF fits, as presented above.
The results can be compared to the combined
results for $\sin^2\theta^{(u,d)}_{\rm eff}$ and $\rho_{(u,d)}$
from the LEP+SLD experiments~\cite{ALEPH:2005ab}:
while the uncertainties are of similar size, the present determinations
consider data from a single experiment only.

A simultaneous determination of \rhopu, \rhopd, \kappu\ and \kappd\ is
performed, i.e.\ a \rhopu+\rhopd+\kappu+\kappd+PDF
fit, and the results are given in the appendix~\ref{appx:results}.
The results are compatible with the SM expectation.
These results exhibit sizeable uncertainties, which are due to the
very strong correlations between the EW parameters. 
The exception is \kappu, which exhibits less strong correlations with the
other EW parameters.

Assuming quark universality ($\rhop{,q}=\rhop{,u}=\rhop{,d}$ and
$\kapp{,q}=\kapp{,u}=\kapp{,d}$),
the results of a \rhop{,q}$+$\kapp{,q}+PDF fit is presented
in table~\ref{tab:rhopzkap} and displayed in figure~\ref{fig:rhopkapp}.
These determinations are dominated by the $u$-type quark couplings.
The $\rhop{,q}$ and $\kapp{,q}$ parameters can be
determined together with the electron parameters $\rhop{,e}$ and
$\kapp{,e}$ in a \rhop{,q}+\kapp{,q}+\rhop{,e}+\kapp{,e}+PDF fit.
Results are given in the appendix~\ref{appx:results} and no significant
deviation from the SM expectation is observed.

Assuming the parameters $\rhop{}$ and $\kapp{}$ to be identical for
quarks and leptons, then denoted as $\rhop{,f}$ and $\kapp{,f}$,
these parameters are determined in a \rhop{,f}$+$\kapp{,f}+PDF fit and
results are again listed in table~\ref{tab:rhopzkap}.
The values exhibit the smallest uncertainties and no significant
deviation from unity is observed as expected in the SM.

\begin{figure}[h]
  \begin{center}
    \includegraphics[width=0.48\textwidth]{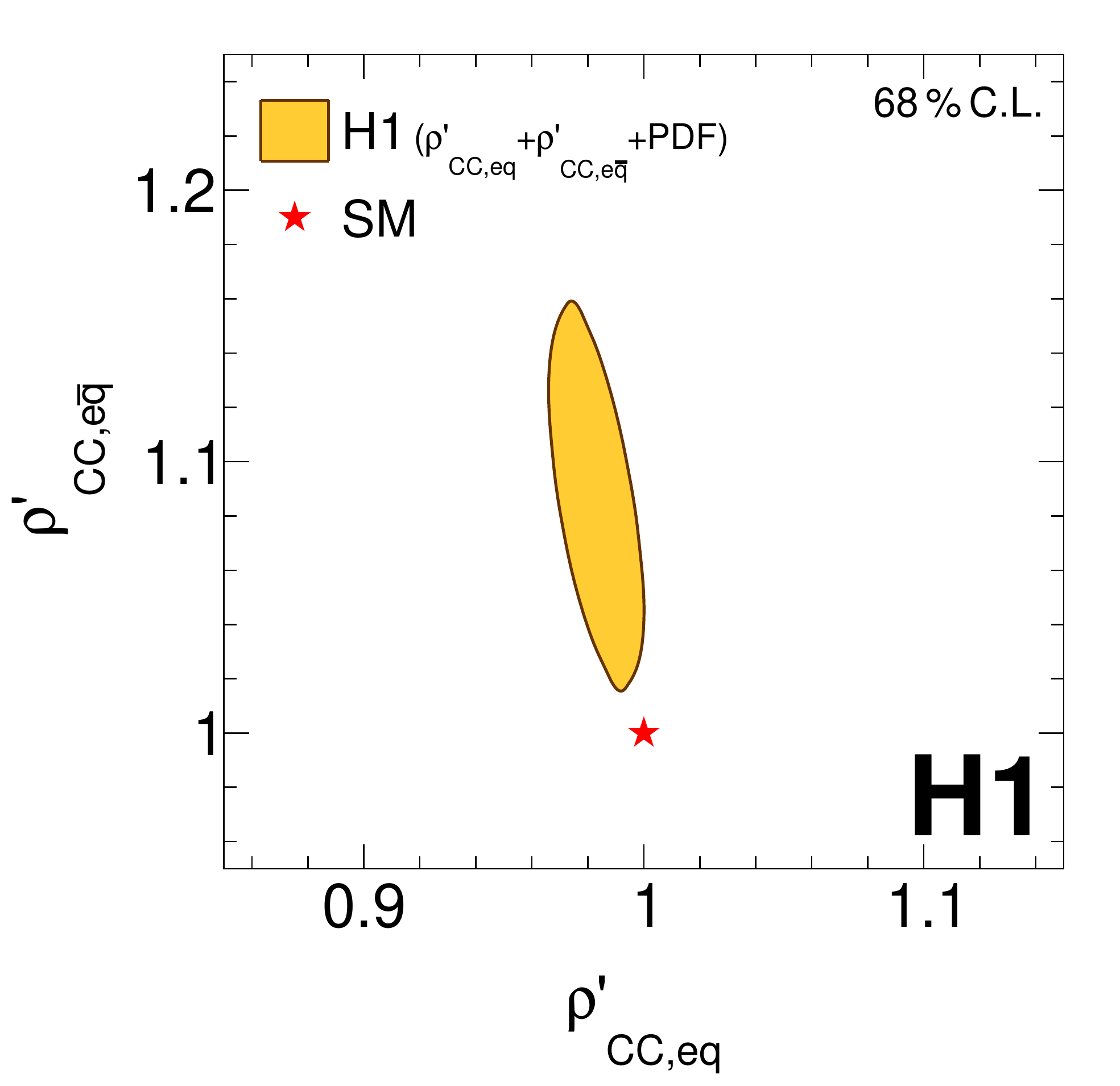
}\hskip0.01\textwidth
  \end{center}
  \caption{
    Results for the \rhopW{,eq}\ and \rhopW{,e\bar{q}} parameters at the
    68\% confidence level (C.L.) obtained with the \rhopW{,eq}+\rhopW{,e\bar{q}}+PDF fit
}
\label{fig:rhopWeqeqbar}
\end{figure}

The values of the $\rhopW{,eq}$ and $\rhopW{,e\bar{q}}$ parameters of the
CC cross sections are determined in a
\rhopW{,eq}$+$\rhopW{,e\bar{q}}+PDF fit and results are 
listed in table~\ref{tab:rhopW}.
The 68\% confidence level contours are shown in figure~\ref{fig:rhopWeqeqbar}.
The parameters are found to be consistent with the SM expectation.
\begin{table}[tb]
  \footnotesize
  \centering
   \begin{tabular}{lccrc}
   \hline
   Fit parameters & \multicolumn{2}{c}{Result} & Correlation \\
   \hline
   \rhopW{,eq}$+$\rhopW{,e\bar{q}}+PDF & $\rhopW{,eq}=0.983\pm0.010$ &$\rhopW{,e\bar{q}}=1.088\pm0.031$ & $-0.50$ \\ %
   \hline
   \rhop{,f}$+$\kapp{,f}$+$\rhopW{,f}+PDF 
      & \multicolumn{2}{c}{see appendix~\ref{appx:results}}  & \\
   \hline
   \end{tabular}
   \caption{
       Results for $\rhopW{}$ parameters. The other parameters $\alpha$, $m_W$, 
       $m_Z$, $m_t$, $m_H$ and $m_f$ are fixed to their SM values.
       The uncertainties quoted correspond to the total uncertainties.
       }
     \label{tab:rhopW}
\end{table}

Setting the two parameters equal, i.e.\
$\rhopW{,f}=\rhopW{,eq}=\rhopW{,e\bar{q}}$, a higher precision is achieved.
The parameter $\rhopW{,f}$ is determined together with the NC parameters
in a \rhop{,f}$+$\kapp{,f}$+$\rhopW{,f}+PDF fit
to $\rhopW{,f}=1.004\pm0.008$.
The full result of that fit is listed in appendix~\ref{appx:results} and
all values are found to be consistent with the SM expectations.
The CC parameter has an uncertainty of 0.8\% and is only weakly
correlated with the NC parameters.
This indicates that the CC and NC parameters can be tested
independently of each other.
The NC parameters are very similar to the ones obtained in the
\rhop{,f}+\kapp{,f}+PDF fit, as presented in table~\ref{tab:rhopzkap}.

The inclusive NC and CC cross sections have been measured over 
a wide range of $\Qsq$ values at HERA.
This can be exploited to perform 
tests of models beyond the SM where scale-dependent modifications 
of coupling parameters are predicted.
Such tests could not be performed by the LEP and SLD
experiments~\cite{Olive:2016xmw}.

In order to study the scale dependence of possible extensions of 
EW parameters in the NC sector the values of \kapp{} and \rhop{} are
determined at different values of \Qsq. 
The data at 
$\Qsq\geq500\,\GeVsq$ are subdivided into four \Qsq ranges
and individual \rhop{} and \kapp{} parameters are assigned to 
each interval. For $\Qsq \leq 500\,\GeVsq$ the SM expectation 
$\rhop{} = 1$ and $\kapp{} = 1$ is used, because of the limited HERA 
sensitivity to EW effects at low energy scales.
All parameters are determined together with a common set of PDF parameters.
Three separate fits are performed:
first, for determining in each $Q^2$ range two quark parameters
\rhop{,q} and \kapp{,q} assuming 
$\rhop{,q}=\rhop{,u}=\rhop{,d}$ and 
$\kapp{,q}=\kapp{,u}=\kapp{,d}$,
while setting the lepton parameters to unity;
second, for determining the lepton parameters \kapp{,e} and \rhop{,e}
while setting the quark parameters to unity;
third, for determining fermion parameters \kapp{,f} and \rhop{,f} common
to both quarks and the lepton assuming $\rhop{,f}=\rhop{,u}=\rhop{,d}=\rhop{,e}$
and $\kapp{,f}=\kapp{,u}=\kapp{,d}=\kapp{,e}$.
Results for the 
\rhop{} and \kapp{} parameters are presented in
figure~\ref{fig:kappQ2} and are given in appendix~\ref{appx:results}.
The values of \rhop{} and \kapp{} in different \Qsq\ intervals 
are largely uncorrelated, while the two parameters \rhop{} and \kapp{} 
within any given \Qsq\ interval have strong correlations.
The highest sensitivity to the \kapp{f} parameter of about 6\% is
found at about $\sqrt{\Qsq}\sim60\,\GeV$.
The results are found to be consistent with the SM expectation and no significant 
scale dependence is observed.
\begin{figure}[tb]
  \begin{center}
    \includegraphics[width=0.48\textwidth]{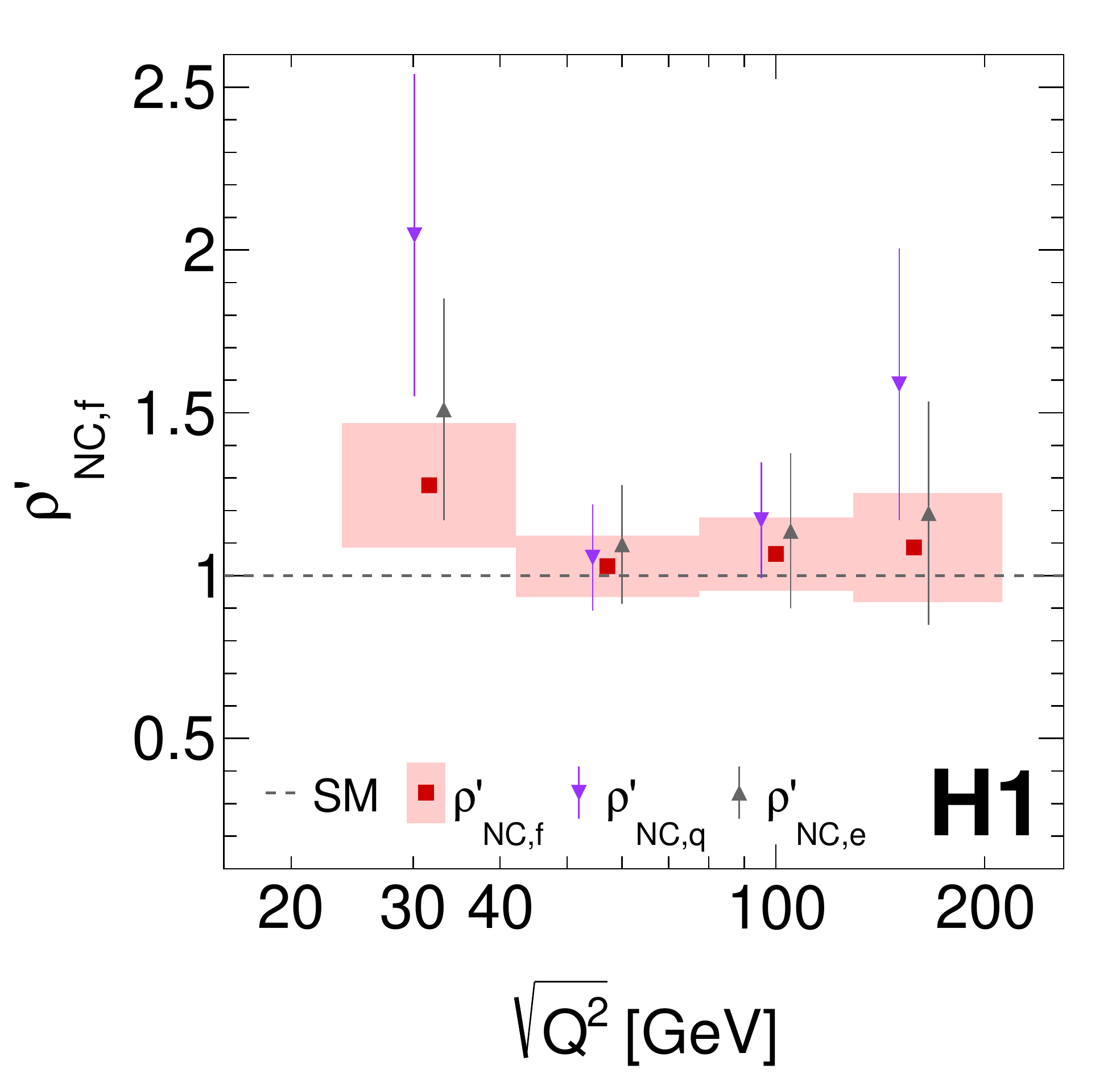%
}\hskip0.01\textwidth
    \includegraphics[width=0.48\textwidth]{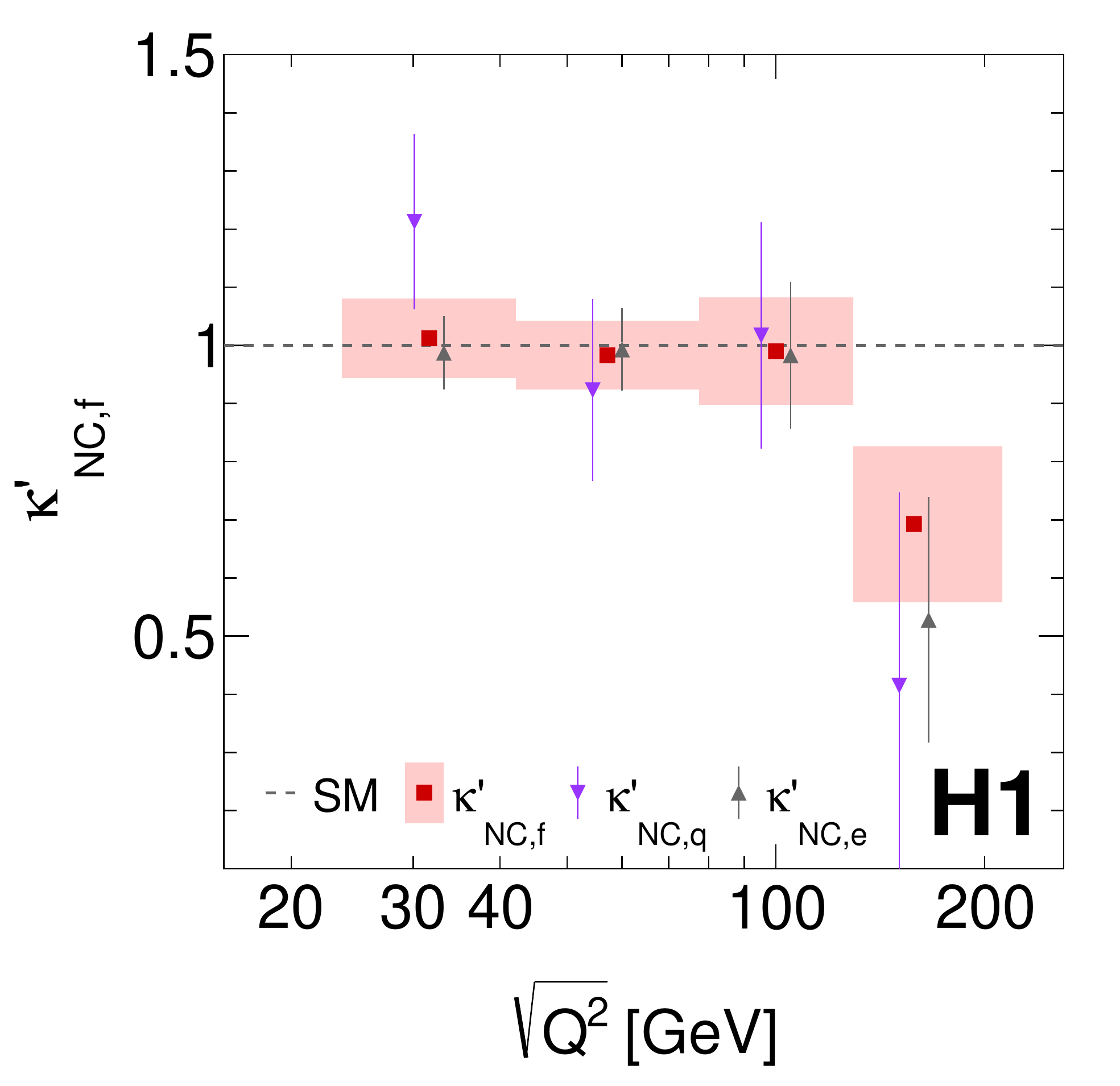}\hskip0.01\textwidth
  \end{center}
  \caption{
    Values of the \rhop{} and \kapp{} parameters determined for
    four different values of \Qsq.
    The error bars, as well as the height of the shaded areas,
    indicate the total uncertainties of the measurement.
    The width of the shaded areas indicates the \Qsq\ range probed by
    the selected data.
    The values for the $\rhop{,q}$, $\rhop{,e}$, $\kapp{,q}$ and $\kapp{,e}$
    parameters are horizontally displaced for better visibility.
}
\label{fig:kappQ2}
\end{figure}

The possible scale dependence of the CC couplings is studied by determining the
$\rhopW{}$ parameters for different values of \Qsq.
A total of three fits are performed, where either $\rhopW{,eq}$
or $\rhopW{,e\bar{q}}$ (c.f.\ equation~\eqref{eq:rhopW}) or $\rhopW{,f}$ is scale dependent.
The CC data are grouped into four $\Qsq$ intervals.
Results of the $\rhopW{}$ parameters are presented in
figure~\ref{fig:rhopW} and are given in the appendix~\ref{appx:results}.
The parameters $\rhopW{,e\bar{q}}$ have uncertainties of
about 4\% over a large range in $\Qsq$, and the parameters $\rhopW{,eq}$ are
determined with a precision of 1.3\% to 3\% over the entire
kinematically accessible range.
The $\rhopW{,f}$ parameters are determined with high
precision of 1.0\% to 1.8\% over the entire $\Qsq$ range.
The values are found to be consistent with the SM expectation of unity.
These studies represent the first determination of the $\rhopW{}$
parameters for separate quark flavours and also its first scale
dependence test.
\begin{figure}[tb]
  \begin{center}
    \includegraphics[width=0.48\textwidth]{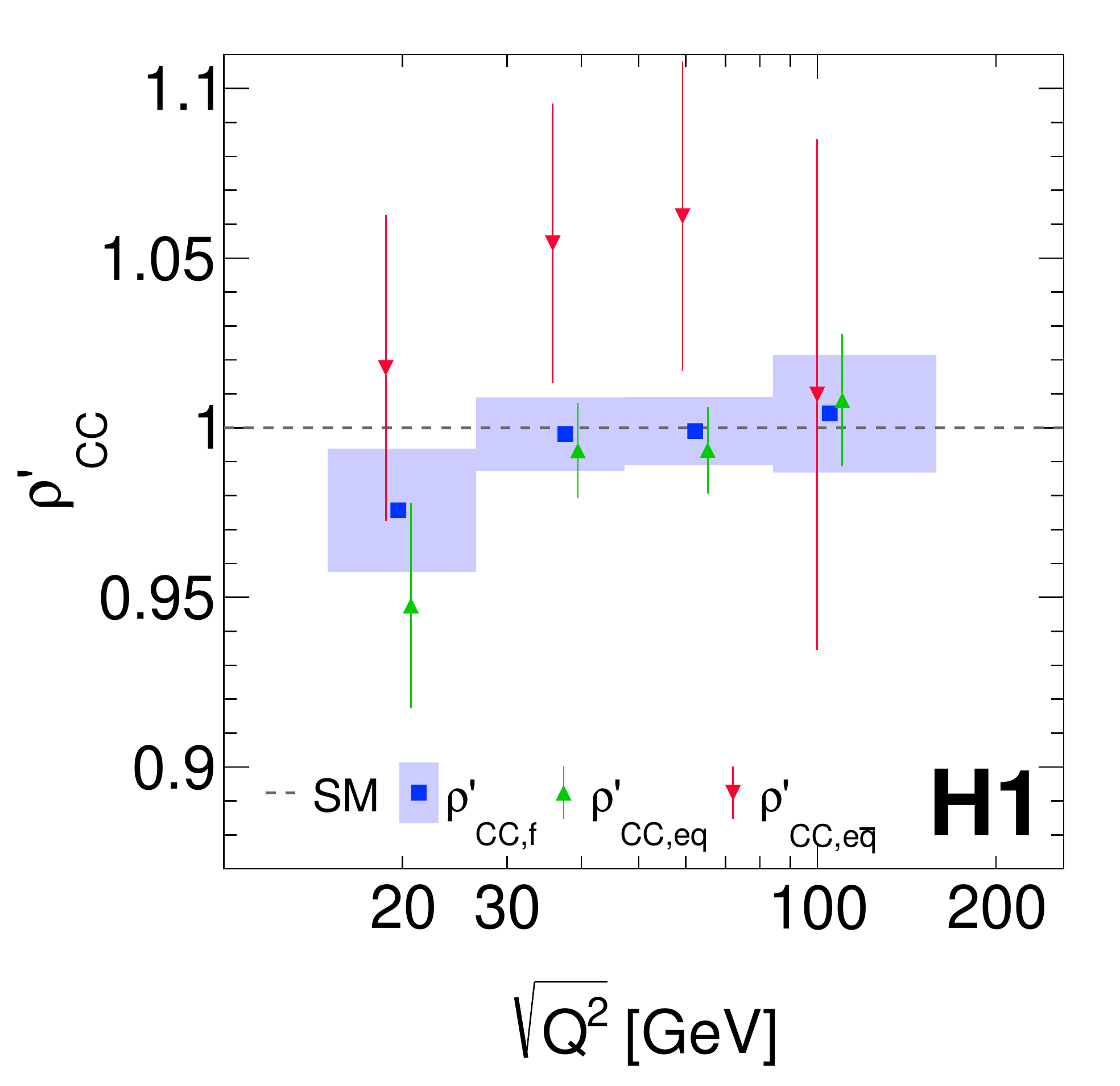
}
    \hskip0.01\textwidth
  \end{center}
  \caption{
    Values of the $\rhopW{}$ parameters determined for
    four different values of $\Qsq$.
    The error bars, as well as the height of the shaded areas,
    indicate the total uncertainties of the measurement.
    The width of the shaded areas indicates the \Qsq\ range probed by
    the selected data.
    The values for the $\rhopW{,eq}$ and $\rhopW{,e\bar{q}}$
    parameters are horizontally displaced for better visibility.
}
\label{fig:rhopW}
\end{figure}

The studies on the scale dependence of the $\rho^{\prime}$ and
$\kappa^{\prime}$ parameters 
provide tests of the SM formalism. Investigations of
specific models beyond the Standard Model such as contact interactions or
leptoquarks, also using the full H1 data sample, have been published
previously \cite{Collaboration:2011qaa,Aaron:2011mv}.

\section{Summary}
\label{sect:Conclusion}
Parameters of the electroweak theory are determined from all 
neutral current and charged current deep-inelastic scattering cross
section measurements published by H1, using NNLO QCD 
and one-loop electroweak predictions.
The inclusion of the cross section data from HERA-II with polarised
lepton beams leads to a substantial improvement in precision  with
respect to the previously published results based on the H1 HERA-I data only.

In combined electroweak and PDF fits, boson and fermion mass parameters 
entering cross section predictions in the on-shell scheme 
are determined simultaneously with the parton distribution functions.
The mass of the $W$ boson is determined from H1 data to
$\mW=80.520\pm 0.115\,\GeV$, fixing \mZ\ to the world average. 
Alternatively the $Z$-boson mass or the top-quark mass are determined with 
uncertainties of $110\,\MeV$ and $26\,\GeV$, respectively, taking
\mW\ to the world average.
Despite their moderate precision, these results are complementary to
direct measurements where particles are produced on-shell in the final
state, since here the mass parameters are determined from purely virtual
particle exchange only.  

The axial-vector and vector weak neutral-current couplings of $u$-
and $d$-type quarks to the $Z$ boson are determined and consistency
with the Standard Model expectation is observed.
The axial-vector and vector couplings of the $u$-type quark are
determined with a precision of about 6\% and 14\%, respectively.

Potential modifications of the weak coupling parameters due to physics beyond the SM are studied
in terms of modifications of the form factors $\rho_\text{NC}$, $\kappa_\text{NC}$ and
$\rho_\text{CC}$.
For this purpose, multiplicative factors to those parameters are introduced, denoted
as \rhop{}, \kapp{}\ and \rhopW{}, respectively.
A precision as good as 7\% or 5\% of the \rhop{,f} and \kapp{,f} parameters
is achieved, respectively.
The $\rhopW{}$ parameters are determined with a precision of up to 8
per mille, and consistency with the Standard Model expectation is
found.
The \Qsq\ dependence of the H1 data allows for a study of the scale
dependence of the $\rhop{}$, $\kapp{}$ and $\rhopW{}$ parameters in the range $12<\sqrt{\Qsq}<100\,\GeV$, 
and no significant deviation from the SM expectation is observed . 
 

\section*{Acknowledgements}
We are grateful to the HERA machine group whose outstanding
efforts have made this experiment possible.
We thank the engineers and technicians for their work in constructing and
maintaining the H1 detector, our funding agencies for
financial support, the
DESY technical staff for continual assistance
and the DESY directorate for support and for the
hospitality which they extend to the non--DESY
members of the collaboration.

We would like to give credit to all partners contributing to the EGI
computing infrastructure for their support for the H1 Collaboration.  

We express our thanks to all those involved in securing not only the
H1 data but also the software and working environment for long term
use, allowing the unique H1 data set to continue to be explored in the
coming years. The transfer from experiment specific to central
resources with long term support, including both storage and batch
systems, has also been crucial to this enterprise. We therefore also
acknowledge the role played by DESY-IT and all people involved during
this transition and their future role in the years to come.



\clearpage
\appendix
\section*{Appendix}

\section{Cross section tables}
\label{appx:cs}
\renewcommand{\arraystretch}{1.0} 
The reduced cross section measurements for NC DIS, as used in this
analysis 
together with their systematic uncertainties \cite{Aaron:2012qi},
for different lepton beam longitudinal 
polarisations and for electron and positron scattering from the
HERA-II running period are given in
tables~\ref{tab:NC1} to~\ref{tab:NC4}, and the differential cross
section for CC DIS are given in tables~\ref{tab:CC1} and~\ref{tab:CC2}.
The reduced cross section is related to the differential cross section,
equation~\eqref{eq:cs}, by
  \begin{equation}
    \sigma_{\rm red} = \frac{d^2\sigma^{\rm NC}}{dxd\Qsq} \frac{xQ^4}{2\pi\alpha^2} \frac{1}{Y_+}\,.
    \label{eq:redcs}
  \end{equation}
The changes compared to the previously published cross
sections~\cite{Aaron:2012qi} comprise the luminosity erratum~\cite{Aaron:2012kn}
and the changes discussed in section~\ref{sec:data}.

\begin{table}[ht]
  \resizebox{\textwidth}{!}{
  \scriptsize
  \centering
  \begin{minipage}{.31\textwidth}
   \begin{tabular}{rclr}
   \hline
\Qsq &  $x$ & $\sigma_{\rm red}$ & $\delta_{\rm stat}$ \\
$[{\GeVsq}]$ &   &  &  $[\%]$ \\
\hline
    120 &  0.0020 &  1.337     &   0.87  \\
    120 &  0.0032 &  1.205     &   1.24  \\
    150 &  0.0032 &  1.218     &   0.73  \\
    150 &  0.0050 &  1.091     &   0.88  \\
    150 &  0.0080 &  0.9375    &   1.20  \\
    150 &  0.0130 &  0.8139    &   1.68  \\
    200 &  0.0032 &  1.247     &   1.35  \\
    200 &  0.0050 &  1.100     &   0.96  \\
    200 &  0.0080 &  0.9576    &   0.99  \\
    200 &  0.0130 &  0.7821    &   1.14  \\
    200 &  0.0200 &  0.6935    &   1.23  \\
    200 &  0.0320 &  0.5849    &   1.38  \\
    200 &  0.0500 &  0.5208    &   1.63  \\
    200 &  0.0800 &  0.4427    &   1.73  \\
    200 &  0.1300 &  0.3591    &   2.09  \\
    200 &  0.1800 &  0.3046    &   2.71  \\
    250 &  0.0050 &  1.118     &   1.12  \\
    250 &  0.0080 &  0.9705    &   1.10  \\
    250 &  0.0130 &  0.8206    &   1.20  \\
    250 &  0.0200 &  0.6944    &   1.23  \\
    250 &  0.0320 &  0.5931    &   1.30  \\
    250 &  0.0500 &  0.5069    &   1.48  \\
    250 &  0.0800 &  0.4251    &   1.52  \\
    250 &  0.1300 &  0.3632    &   1.54  \\
    250 &  0.1800 &  0.3097    &   2.11  \\
    300 &  0.0050 &  1.133     &   1.89  \\
    300 &  0.0080 &  0.9826    &   1.28  \\
    300 &  0.0130 &  0.8196    &   1.28  \\
    300 &  0.0200 &  0.7027    &   1.42  \\
    300 &  0.0320 &  0.5867    &   1.50  \\
    300 &  0.0500 &  0.4994    &   1.62  \\
    300 &  0.0800 &  0.4250    &   1.72  \\
    300 &  0.1300 &  0.3621    &   1.71  \\
    300 &  0.1800 &  0.3023    &   2.26  \\
    300 &  0.4000 &  0.1468    &   2.75  \\
    400 &  0.0080 &  1.048     &   1.54  \\
    400 &  0.0130 &  0.8622    &   1.50  \\
    400 &  0.0200 &  0.7260    &   1.54  \\
    400 &  0.0320 &  0.6114    &   1.63  \\
    400 &  0.0500 &  0.4951    &   1.84  \\
    400 &  0.0800 &  0.4279    &   1.91  \\
    400 &  0.1300 &  0.3676    &   1.93  \\
    400 &  0.1800 &  0.3055    &   2.43  \\
    400 &  0.4000 &  0.1469    &   3.09  \\
   \hline
\\
\\
\\
\\
\\
\\
   \end{tabular}
  \end{minipage}
  \hskip.02\textwidth
  \begin{minipage}{.31\textwidth}
    \begin{tabular}{rclr}
\hline
      \Qsq &  $x$ & $\sigma_{\rm red}$ & $\delta_{\rm stat}$ \\
$[{\GeVsq}]$ &   &  &  $[\%]$ \\
\hline
    500 &  0.0080 &  1.010     &   2.57  \\
    500 &  0.0130 &  0.9106    &   1.85  \\
    500 &  0.0200 &  0.7435    &   1.83  \\
    500 &  0.0320 &  0.6373    &   1.87  \\
    500 &  0.0500 &  0.5533    &   1.99  \\
    500 &  0.0800 &  0.4263    &   2.27  \\
    500 &  0.1300 &  0.3740    &   2.54  \\
    500 &  0.1800 &  0.3373    &   2.86  \\
    500 &  0.2500 &  0.2585    &   3.32  \\
    650 &  0.0130 &  0.9046    &   2.08  \\
    650 &  0.0200 &  0.7765    &   2.14  \\
    650 &  0.0320 &  0.6486    &   2.23  \\
    650 &  0.0500 &  0.5354    &   2.35  \\
    650 &  0.0800 &  0.4403    &   2.66  \\
    650 &  0.1300 &  0.3684    &   2.94  \\
    650 &  0.1800 &  0.3215    &   3.18  \\
    650 &  0.2500 &  0.2529    &   4.13  \\
    650 &  0.4000 &  0.1251    &   6.14  \\
    800 &  0.0130 &  0.9258    &   3.50  \\
    800 &  0.0200 &  0.7391    &   2.51  \\
    800 &  0.0320 &  0.6353    &   2.67  \\
    800 &  0.0500 &  0.5523    &   2.74  \\
    800 &  0.0800 &  0.4430    &   3.04  \\
    800 &  0.1300 &  0.3476    &   3.58  \\
    800 &  0.1800 &  0.3205    &   3.75  \\
    800 &  0.2500 &  0.2468    &   4.63  \\
    800 &  0.4000 &  0.1373    &   5.93  \\
   1000 &  0.0130 &  0.8664    &   3.45  \\
   1000 &  0.0200 &  0.7899    &   2.87  \\
   1000 &  0.0320 &  0.6760    &   2.82  \\
   1000 &  0.0500 &  0.5166    &   3.15  \\
   1000 &  0.0800 &  0.4428    &   3.43  \\
   1000 &  0.1300 &  0.3396    &   4.21  \\
   1000 &  0.1800 &  0.3682    &   3.98  \\
   1000 &  0.2500 &  0.2659    &   4.61  \\
   1000 &  0.4000 &  0.1299    &   6.56  \\
   1200 &  0.0130 &  0.9440    &   5.43  \\
   1200 &  0.0200 &  0.7891    &   3.60  \\
   1200 &  0.0320 &  0.6964    &   3.27  \\
   1200 &  0.0500 &  0.5465    &   3.48  \\
   1200 &  0.0800 &  0.4591    &   3.73  \\
   1200 &  0.1300 &  0.3602    &   5.38  \\
   1200 &  0.1800 &  0.3308    &   4.65  \\
   1200 &  0.2500 &  0.2207    &   5.58  \\
   1200 &  0.4000 &  0.1264    &   7.08  \\
\hline
\\
\\
\\
\\
\\
    \end{tabular}
  \end{minipage}
  \hskip.02\textwidth
  \begin{minipage}{.31\textwidth}
    \begin{tabular}{rclr}
\hline
      \Qsq &  $x$ & $\sigma_{\rm red}$  & $\delta_{\rm stat}$ \\
$[{\GeVsq}]$ &   &  &  $[\%]$ \\
\hline
   1500 &  0.0200 &  0.8335    &   4.26  \\
   1500 &  0.0320 &  0.6943    &   4.06  \\
   1500 &  0.0500 &  0.5646    &   4.02  \\
   1500 &  0.0800 &  0.5143    &   4.05  \\
   1500 &  0.1300 &  0.3622    &   5.25  \\
   1500 &  0.1800 &  0.3159    &   5.42  \\
   1500 &  0.2500 &  0.2365    &   6.05  \\
   1500 &  0.4000 &  0.1393    &   8.82  \\
   1500 &  0.6500 &  0.01511   &  14.78  \\
   2000 &  0.0219 &  0.9308    &   6.58  \\
   2000 &  0.0320 &  0.6562    &   4.89  \\
   2000 &  0.0500 &  0.5678    &   4.87  \\
   2000 &  0.0800 &  0.4520    &   5.02  \\
   2000 &  0.1300 &  0.3780    &   5.98  \\
   2000 &  0.1800 &  0.3071    &   6.52  \\
   2000 &  0.2500 &  0.2566    &   6.68  \\
   2000 &  0.4000 &  0.1289    &   8.56  \\
   2000 &  0.6500 &  0.01095   &  19.67  \\
   3000 &  0.0320 &  0.8036    &   4.41  \\
   3000 &  0.0500 &  0.6145    &   4.01  \\
   3000 &  0.0800 &  0.5119    &   4.37  \\
   3000 &  0.1300 &  0.4313    &   5.17  \\
   3000 &  0.1800 &  0.3004    &   6.14  \\
   3000 &  0.2500 &  0.2216    &   6.55  \\
   3000 &  0.4000 &  0.1292    &   7.49  \\
   3000 &  0.6500 &  0.01350   &  14.62  \\
   5000 &  0.0547 &  0.6974    &   5.98  \\
   5000 &  0.0800 &  0.5881    &   4.65  \\
   5000 &  0.1300 &  0.5103    &   5.23  \\
   5000 &  0.1800 &  0.3976    &   6.13  \\
   5000 &  0.2500 &  0.2348    &   8.02  \\
   5000 &  0.4000 &  0.1101    &   9.88  \\
   5000 &  0.6500 &  0.01502   &  16.48  \\
   8000 &  0.0875 &  0.6943    &   8.89  \\
   8000 &  0.1300 &  0.5661    &   7.10  \\
   8000 &  0.1800 &  0.4017    &   8.01  \\
   8000 &  0.2500 &  0.2807    &   9.07  \\
   8000 &  0.4000 &  0.1232    &  12.62  \\
   8000 &  0.6500 &  0.01091   &  21.89  \\
  12000 &  0.1300 &  0.7921    &  15.45  \\
  12000 &  0.1800 &  0.5805    &   9.59  \\
  12000 &  0.2500 &  0.3347    &  11.15  \\
  12000 &  0.4000 &  0.2244    &  12.42  \\
  12000 &  0.6500 &  0.01526   &  27.80  \\
  20000 &  0.2500 &  0.6549    &  13.34  \\
  20000 &  0.4000 &  0.2329    &  16.55  \\
  20000 &  0.6500 &  0.01985   &  40.89  \\
  30000 &  0.4000 &  0.1845    &  36.01  \\
  30000 &  0.6500 &  0.04510   &  37.83  \\
  50000 &  0.6500 &  0.1250    &  57.78  \\
\hline
    \end{tabular}
  \end{minipage}
} 
   \caption{
     The NC $e^-p$ reduced cross section $\sigma_{\rm red}$ with lepton beam
     polarisation $P_e=-25.8\%$ with their 
     statistical ($\delta_{\rm stat}$) uncertainties.
     The full uncertainties are available in
     ref.~\cite{Aaron:2012qi}, while the respective cross section values
     are updated according to section~\ref{sec:data} and ref.~\cite{Aaron:2012kn}.
   }
   \label{tab:NC1}
\end{table}

\begin{table}[ht]
  \resizebox{\textwidth}{!}{
  \scriptsize
  \centering
  \begin{minipage}{.31\textwidth}
   \begin{tabular}{rclr}
   \hline
\Qsq &  $x$ & $\sigma_{\rm red}$  & $\delta_{\rm stat}$ \\
$[{\GeVsq}]$ &   &  & $[\%]$ \\
\hline
    120 &  0.0020 &  1.340     &   1.29  \\
    120 &  0.0032 &  1.213     &   1.78  \\
    150 &  0.0032 &  1.208     &   1.09  \\
    150 &  0.0050 &  1.104     &   1.29  \\
    150 &  0.0080 &  0.9534    &   1.78  \\
    150 &  0.0130 &  0.7840    &   2.42  \\
    200 &  0.0032 &  1.189     &   2.06  \\
    200 &  0.0050 &  1.092     &   1.46  \\
    200 &  0.0080 &  0.9487    &   1.44  \\
    200 &  0.0130 &  0.7938    &   1.67  \\
    200 &  0.0200 &  0.6910    &   1.81  \\
    200 &  0.0320 &  0.5630    &   2.12  \\
    200 &  0.0500 &  0.5323    &   2.48  \\
    200 &  0.0800 &  0.4308    &   2.51  \\
    200 &  0.1300 &  0.3616    &   2.84  \\
    200 &  0.1800 &  0.3113    &   4.12  \\
    250 &  0.0050 &  1.100     &   1.69  \\
    250 &  0.0080 &  0.9277    &   1.64  \\
    250 &  0.0130 &  0.7978    &   1.80  \\
    250 &  0.0200 &  0.6690    &   1.86  \\
    250 &  0.0320 &  0.5657    &   1.95  \\
    250 &  0.0500 &  0.4677    &   2.20  \\
    250 &  0.0800 &  0.4305    &   2.24  \\
    250 &  0.1300 &  0.3710    &   2.29  \\
    250 &  0.1800 &  0.3035    &   3.24  \\
    300 &  0.0050 &  1.163     &   2.77  \\
    300 &  0.0080 &  0.9754    &   1.89  \\
    300 &  0.0130 &  0.8091    &   1.92  \\
    300 &  0.0200 &  0.6930    &   2.10  \\
    300 &  0.0320 &  0.5937    &   2.18  \\
    300 &  0.0500 &  0.5014    &   2.46  \\
    300 &  0.0800 &  0.4269    &   2.56  \\
    300 &  0.1300 &  0.3530    &   2.61  \\
    300 &  0.1800 &  0.2847    &   3.47  \\
    300 &  0.4000 &  0.1523    &   3.91  \\
    400 &  0.0080 &  0.9979    &   2.41  \\
    400 &  0.0130 &  0.8314    &   2.24  \\
    400 &  0.0200 &  0.6742    &   2.36  \\
    400 &  0.0320 &  0.5909    &   2.46  \\
    400 &  0.0500 &  0.4953    &   2.70  \\
    400 &  0.0800 &  0.3995    &   3.05  \\
    400 &  0.1300 &  0.3666    &   2.96  \\
    400 &  0.1800 &  0.3074    &   3.45  \\
    400 &  0.4000 &  0.1482    &   4.99  \\
\hline
\\
\\
\\
\\
\\
    \end{tabular}
  \end{minipage}
  \hskip.02\textwidth
  \begin{minipage}{.31\textwidth}
    \begin{tabular}{rclr}
\hline
      \Qsq &  $x$ & $\sigma_{\rm red}$ & $\delta_{\rm stat}$ \\
$[{\GeVsq}]$ &   &  &  $[\%]$ \\
\hline
    500 &  0.0080 &  0.9586    &   3.93  \\
    500 &  0.0130 &  0.8227    &   2.80  \\
    500 &  0.0200 &  0.6873    &   2.80  \\
    500 &  0.0320 &  0.5849    &   2.89  \\
    500 &  0.0500 &  0.5161    &   3.01  \\
    500 &  0.0800 &  0.4334    &   3.28  \\
    500 &  0.1300 &  0.3687    &   4.14  \\
    500 &  0.1800 &  0.3218    &   4.06  \\
    500 &  0.2500 &  0.2447    &   5.05  \\
    650 &  0.0130 &  0.8753    &   3.13  \\
    650 &  0.0200 &  0.7334    &   3.26  \\
    650 &  0.0320 &  0.6383    &   3.33  \\
    650 &  0.0500 &  0.5511    &   3.46  \\
    650 &  0.0800 &  0.4102    &   4.01  \\
    650 &  0.1300 &  0.3354    &   4.99  \\
    650 &  0.1800 &  0.3324    &   4.67  \\
    650 &  0.2500 &  0.2521    &   5.55  \\
    650 &  0.4000 &  0.1130    &   8.49  \\
    800 &  0.0130 &  0.8344    &   5.20  \\
    800 &  0.0200 &  0.7130    &   3.76  \\
    800 &  0.0320 &  0.6115    &   3.87  \\
    800 &  0.0500 &  0.5470    &   4.04  \\
    800 &  0.0800 &  0.3842    &   4.83  \\
    800 &  0.1300 &  0.3592    &   5.90  \\
    800 &  0.1800 &  0.3187    &   6.28  \\
    800 &  0.2500 &  0.2272    &   6.66  \\
    800 &  0.4000 &  0.1210    &   9.46  \\
   1000 &  0.0130 &  0.8399    &   5.19  \\
   1000 &  0.0200 &  0.7135    &   4.48  \\
   1000 &  0.0320 &  0.6349    &   4.66  \\
   1000 &  0.0500 &  0.5027    &   4.74  \\
   1000 &  0.0800 &  0.4182    &   5.21  \\
   1000 &  0.1300 &  0.3902    &   5.82  \\
   1000 &  0.1800 &  0.3002    &   6.43  \\
   1000 &  0.2500 &  0.2774    &   6.71  \\
   1000 &  0.4000 &  0.1267    &   9.92  \\
   1200 &  0.0130 &  0.7777    &   9.00  \\
   1200 &  0.0200 &  0.7689    &   5.37  \\
   1200 &  0.0320 &  0.6439    &   5.03  \\
   1200 &  0.0500 &  0.5285    &   5.22  \\
   1200 &  0.0800 &  0.4649    &   5.53  \\
   1200 &  0.1300 &  0.3395    &   7.00  \\
   1200 &  0.1800 &  0.2714    &   7.60  \\
   1200 &  0.2500 &  0.2206    &   8.26  \\
   1200 &  0.4000 &  0.1337    &  10.01  \\
\hline
\\
\\
\\
\\
    \end{tabular}
  \end{minipage}
  \hskip.02\textwidth
  \begin{minipage}{.31\textwidth}
    \begin{tabular}{rclr}
\hline
      \Qsq &  $x$ & $\sigma_{\rm red}$ & $\delta_{\rm stat}$ \\
$[{\GeVsq}]$ &   &  &  $[\%]$ \\
\hline
   1500 &  0.0200 &  0.7317    &   6.68  \\
   1500 &  0.0320 &  0.6439    &   6.22  \\
   1500 &  0.0500 &  0.5514    &   6.00  \\
   1500 &  0.0800 &  0.4600    &   6.35  \\
   1500 &  0.1300 &  0.3344    &  10.17  \\
   1500 &  0.1800 &  0.2695    &   8.79  \\
   1500 &  0.2500 &  0.2555    &   8.68  \\
   1500 &  0.4000 &  0.09316   &  13.16  \\
   1500 &  0.6500 &  0.01262   &  23.63  \\
   2000 &  0.0219 &  0.7628    &  10.62  \\
   2000 &  0.0320 &  0.6464    &   7.29  \\
   2000 &  0.0500 &  0.5190    &   7.57  \\
   2000 &  0.0800 &  0.4552    &   7.37  \\
   2000 &  0.1300 &  0.3166    &   9.69  \\
   2000 &  0.1800 &  0.2939    &   9.83  \\
   2000 &  0.2500 &  0.2322    &  10.39  \\
   2000 &  0.4000 &  0.1216    &  12.93  \\
   2000 &  0.6500 &  0.008022  &  33.44  \\
   3000 &  0.0320 &  0.6126    &   7.46  \\
   3000 &  0.0500 &  0.6022    &   5.96  \\
   3000 &  0.0800 &  0.4925    &   6.63  \\
   3000 &  0.1300 &  0.3542    &   8.44  \\
   3000 &  0.1800 &  0.3105    &   9.00  \\
   3000 &  0.2500 &  0.2919    &   8.59  \\
   3000 &  0.4000 &  0.09196   &  12.93  \\
   3000 &  0.6500 &  0.005166  &  35.57  \\
   5000 &  0.0547 &  0.5881    &   9.54  \\
   5000 &  0.0800 &  0.4575    &   7.68  \\
   5000 &  0.1300 &  0.4144    &   8.49  \\
   5000 &  0.1800 &  0.3602    &   9.47  \\
   5000 &  0.2500 &  0.2529    &  16.81  \\
   5000 &  0.4000 &  0.1434    &  12.82  \\
   5000 &  0.6500 &  0.01324   &  25.88  \\
   8000 &  0.0875 &  0.6279    &  13.94  \\
   8000 &  0.1300 &  0.4992    &  11.13  \\
   8000 &  0.1800 &  0.3997    &  11.74  \\
   8000 &  0.2500 &  0.2553    &  14.04  \\
   8000 &  0.4000 &  0.1182    &  18.93  \\
   8000 &  0.6500 &  0.01682   &  26.77  \\
  12000 &  0.1300 &  0.7385    &  23.42  \\
  12000 &  0.1800 &  0.4153    &  16.73  \\
  12000 &  0.2500 &  0.3198    &  16.72  \\
  12000 &  0.4000 &  0.1575    &  21.86  \\
  12000 &  0.6500 &  0.01281   &  44.83  \\
  20000 &  0.2500 &  0.2146    &  34.11  \\
  20000 &  0.4000 &  0.2378    &  24.30  \\
  20000 &  0.6500 &  0.01372   &  70.89  \\
  30000 &  0.4000 &  0.2765    &  43.40  \\
  30000 &  0.6500 &  0.04110   &  57.81  \\
\hline
    \end{tabular}
  \end{minipage}
} 
   \caption{
     The NC $e^-p$ reduced cross section $\sigma_{\rm red}$ with lepton beam
     polarisation $P_e=36.0\%$ with their 
     statistical ($\delta_{\rm stat}$) uncertainties.
     The full uncertainties are available in
     ref.~\cite{Aaron:2012qi}, while the respective cross section values
     are updated according to section~\ref{sec:data} and ref.~\cite{Aaron:2012kn}.
   }
   \label{tab:NC2}
\end{table}

\begin{table}[ht]
  \resizebox{\textwidth}{!}{
  \scriptsize
  \centering
  \begin{minipage}{.31\textwidth}
   \begin{tabular}{rclr}
   \hline
\Qsq &  $x$ & $\sigma_{\rm red}$ & $\delta_{\rm stat}$ \\
$[{\GeVsq}]$ &   &  &  $[\%]$ \\
\hline
    120 &  0.0020 &  1.367     &   0.97  \\
    120 &  0.0032 &  1.249     &   1.38  \\
    150 &  0.0032 &  1.248     &   0.82  \\
    150 &  0.0050 &  1.096     &   1.00  \\
    150 &  0.0080 &  0.9470    &   1.36  \\
    150 &  0.0130 &  0.8224    &   1.92  \\
    200 &  0.0032 &  1.263     &   1.55  \\
    200 &  0.0050 &  1.122     &   1.08  \\
    200 &  0.0080 &  0.9667    &   1.09  \\
    200 &  0.0130 &  0.8071    &   1.24  \\
    200 &  0.0200 &  0.7003    &   1.38  \\
    200 &  0.0320 &  0.5918    &   1.61  \\
    200 &  0.0500 &  0.5312    &   1.79  \\
    200 &  0.0800 &  0.4385    &   1.99  \\
    200 &  0.1300 &  0.3722    &   2.25  \\
    200 &  0.1800 &  0.3266    &   2.97  \\
    250 &  0.0050 &  1.128     &   1.25  \\
    250 &  0.0080 &  0.9659    &   1.24  \\
    250 &  0.0130 &  0.8085    &   1.38  \\
    250 &  0.0200 &  0.6896    &   1.40  \\
    250 &  0.0320 &  0.5789    &   1.46  \\
    250 &  0.0500 &  0.5079    &   1.57  \\
    250 &  0.0800 &  0.4438    &   1.71  \\
    250 &  0.1300 &  0.3836    &   1.81  \\
    250 &  0.1800 &  0.3011    &   2.39  \\
    300 &  0.0050 &  1.135     &   2.13  \\
    300 &  0.0080 &  0.9749    &   1.45  \\
    300 &  0.0130 &  0.8181    &   1.45  \\
    300 &  0.0200 &  0.7086    &   1.60  \\
    300 &  0.0320 &  0.5926    &   1.69  \\
    300 &  0.0500 &  0.5053    &   1.82  \\
    300 &  0.0800 &  0.4462    &   1.85  \\
    300 &  0.1300 &  0.3717    &   1.93  \\
    300 &  0.1800 &  0.3081    &   2.52  \\
    300 &  0.4000 &  0.1551    &   3.06  \\
    400 &  0.0080 &  1.025     &   1.77  \\
    400 &  0.0130 &  0.8345    &   1.71  \\
    400 &  0.0200 &  0.7131    &   1.77  \\
    400 &  0.0320 &  0.6080    &   1.91  \\
    400 &  0.0500 &  0.5019    &   2.05  \\
    400 &  0.0800 &  0.4265    &   2.15  \\
    400 &  0.1300 &  0.3662    &   2.11  \\
    400 &  0.1800 &  0.3066    &   2.76  \\
    400 &  0.4000 &  0.1572    &   3.56  \\
\hline
\\
\\
\\
\\
    \end{tabular}
  \end{minipage}
  \hskip.02\textwidth
  \begin{minipage}{.31\textwidth}
    \begin{tabular}{rclr}
\hline
      \Qsq &  $x$ & $\sigma_{\rm red}$ & $\delta_{\rm stat}$ \\
$[{\GeVsq}]$ &   &  &  $[\%]$ \\
\hline
    500 &  0.0080 &  0.9862    &   2.93  \\
    500 &  0.0130 &  0.8805    &   2.10  \\
    500 &  0.0200 &  0.7446    &   2.09  \\
    500 &  0.0320 &  0.6097    &   2.28  \\
    500 &  0.0500 &  0.5252    &   2.29  \\
    500 &  0.0800 &  0.4306    &   2.48  \\
    500 &  0.1300 &  0.4018    &   2.93  \\
    500 &  0.1800 &  0.3160    &   3.13  \\
    500 &  0.2500 &  0.2502    &   3.83  \\
    650 &  0.0130 &  0.8789    &   2.35  \\
    650 &  0.0200 &  0.7456    &   2.47  \\
    650 &  0.0320 &  0.6240    &   2.56  \\
    650 &  0.0500 &  0.5102    &   2.74  \\
    650 &  0.0800 &  0.4037    &   3.04  \\
    650 &  0.1300 &  0.3624    &   3.30  \\
    650 &  0.1800 &  0.3269    &   3.57  \\
    650 &  0.2500 &  0.2449    &   4.65  \\
    650 &  0.4000 &  0.1366    &   6.70  \\
    800 &  0.0130 &  0.7990    &   3.94  \\
    800 &  0.0200 &  0.7034    &   2.84  \\
    800 &  0.0320 &  0.5953    &   3.09  \\
    800 &  0.0500 &  0.5276    &   3.14  \\
    800 &  0.0800 &  0.4697    &   3.35  \\
    800 &  0.1300 &  0.3511    &   4.04  \\
    800 &  0.1800 &  0.3237    &   4.18  \\
    800 &  0.2500 &  0.2226    &   5.17  \\
    800 &  0.4000 &  0.1247    &   7.21  \\
   1000 &  0.0130 &  0.8326    &   3.89  \\
   1000 &  0.0200 &  0.7443    &   3.28  \\
   1000 &  0.0320 &  0.5882    &   3.38  \\
   1000 &  0.0500 &  0.5003    &   3.58  \\
   1000 &  0.0800 &  0.4275    &   3.88  \\
   1000 &  0.1300 &  0.3378    &   4.75  \\
   1000 &  0.1800 &  0.3008    &   4.92  \\
   1000 &  0.2500 &  0.2354    &   5.56  \\
   1000 &  0.4000 &  0.1210    &   7.75  \\
   1200 &  0.0130 &  0.7975    &   6.75  \\
   1200 &  0.0200 &  0.6749    &   4.32  \\
   1200 &  0.0320 &  0.6406    &   3.76  \\
   1200 &  0.0500 &  0.5253    &   3.95  \\
   1200 &  0.0800 &  0.4256    &   4.33  \\
   1200 &  0.1300 &  0.3242    &   5.38  \\
   1200 &  0.1800 &  0.2971    &   5.47  \\
   1200 &  0.2500 &  0.2680    &   5.62  \\
   1200 &  0.4000 &  0.1086    &   8.59  \\
\hline
\\
\\
\\
    \end{tabular}
  \end{minipage}
  \hskip.02\textwidth
  \begin{minipage}{.31\textwidth}
    \begin{tabular}{rclr}
\hline
      \Qsq &  $x$ & $\sigma_{\rm red}$ & $\delta_{\rm stat}$ \\
$[{\GeVsq}]$ &   &  &  $[\%]$ \\
\hline
   1500 &  0.0200 &  0.6695    &   5.31  \\
   1500 &  0.0320 &  0.5980    &   5.22  \\
   1500 &  0.0500 &  0.5295    &   4.55  \\
   1500 &  0.0800 &  0.4702    &   5.12  \\
   1500 &  0.1300 &  0.3057    &   6.35  \\
   1500 &  0.1800 &  0.2927    &   6.32  \\
   1500 &  0.2500 &  0.2585    &   6.39  \\
   1500 &  0.4000 &  0.1211    &   8.89  \\
   1500 &  0.6500 &  0.01573   &  16.04  \\
   2000 &  0.0219 &  0.6690    &   8.55  \\
   2000 &  0.0320 &  0.5502    &   5.99  \\
   2000 &  0.0500 &  0.5168    &   5.63  \\
   2000 &  0.0800 &  0.4365    &   5.55  \\
   2000 &  0.1300 &  0.3138    &   7.22  \\
   2000 &  0.1800 &  0.2954    &   7.37  \\
   2000 &  0.2500 &  0.2150    &   7.85  \\
   2000 &  0.4000 &  0.1188    &   9.92  \\
   2000 &  0.6500 &  0.01324   &  19.28  \\
   3000 &  0.0320 &  0.5883    &   5.61  \\
   3000 &  0.0500 &  0.4774    &   5.02  \\
   3000 &  0.0800 &  0.4114    &   5.36  \\
   3000 &  0.1300 &  0.3340    &   6.38  \\
   3000 &  0.1800 &  0.2711    &   7.11  \\
   3000 &  0.2500 &  0.2219    &   7.07  \\
   3000 &  0.4000 &  0.1272    &   8.29  \\
   3000 &  0.6500 &  0.01302   &  16.94  \\
   5000 &  0.0547 &  0.4324    &   7.86  \\
   5000 &  0.0800 &  0.3520    &   6.38  \\
   5000 &  0.1300 &  0.3150    &   7.32  \\
   5000 &  0.1800 &  0.2647    &   8.19  \\
   5000 &  0.2500 &  0.2278    &   8.92  \\
   5000 &  0.4000 &  0.09719   &  11.65  \\
   5000 &  0.6500 &  0.007011  &  27.87  \\
   8000 &  0.0875 &  0.2552    &  15.58  \\
   8000 &  0.1300 &  0.2586    &  11.14  \\
   8000 &  0.1800 &  0.2346    &  11.31  \\
   8000 &  0.2500 &  0.2234    &  11.01  \\
   8000 &  0.4000 &  0.1034    &  15.10  \\
   8000 &  0.6500 &  0.01192   &  25.07  \\
  12000 &  0.1300 &  0.2033    &  28.52  \\
  12000 &  0.1800 &  0.2078    &  17.53  \\
  12000 &  0.2500 &  0.1426    &  18.70  \\
  12000 &  0.4000 &  0.07284   &  24.35  \\
  12000 &  0.6500 &  0.008088  &  44.93  \\
  20000 &  0.2500 &  0.1039    &  32.86  \\
  20000 &  0.4000 &  0.07670   &  31.82  \\
  20000 &  0.6500 &  0.01353   &  57.87  \\
\hline
  \\
    \end{tabular}
  \end{minipage}
} 
 \caption{
     The NC $e^+p$ reduced cross section $\sigma_{\rm red}$ with lepton beam
     polarisation $P_e=-37.0\%$ with their  
     statistical ($\delta_{\rm stat}$) uncertainties.
     The full uncertainties are available in
     ref.~\cite{Aaron:2012qi}, while the respective cross section values
     are updated according to section~\ref{sec:data} and ref.~\cite{Aaron:2012kn}.
   }
 \label{tab:NC3}
\end{table}

\begin{table}[ht]
  \resizebox{\textwidth}{!}{
    \scriptsize
  \centering
  \begin{minipage}{.31\textwidth}
   \begin{tabular}{rclr}
   \hline
   \Qsq &  $x$ & $\sigma_{\rm red}$ & $\delta_{\rm stat}$ \\
$[{\GeVsq}]$ &   &  &  $[\%]$ \\
   \hline
    120 &  0.0020 &  1.353     &   0.87  \\
    120 &  0.0032 &  1.192     &   1.27  \\
    150 &  0.0032 &  1.224     &   0.74  \\
    150 &  0.0050 &  1.096     &   0.88  \\
    150 &  0.0080 &  0.9530    &   1.22  \\
    150 &  0.0130 &  0.7836    &   1.71  \\
    200 &  0.0032 &  1.225     &   1.40  \\
    200 &  0.0050 &  1.094     &   0.97  \\
    200 &  0.0080 &  0.9510    &   0.98  \\
    200 &  0.0130 &  0.7985    &   1.11  \\
    200 &  0.0200 &  0.6889    &   1.22  \\
    200 &  0.0320 &  0.5832    &   1.40  \\
    200 &  0.0500 &  0.5022    &   1.62  \\
    200 &  0.0800 &  0.4385    &   1.77  \\
    200 &  0.1300 &  0.3558    &   1.96  \\
    200 &  0.1800 &  0.3053    &   2.68  \\
    250 &  0.0050 &  1.124     &   1.13  \\
    250 &  0.0080 &  0.9603    &   1.10  \\
    250 &  0.0130 &  0.8134    &   1.22  \\
    250 &  0.0200 &  0.7022    &   1.25  \\
    250 &  0.0320 &  0.5830    &   1.31  \\
    250 &  0.0500 &  0.5018    &   1.45  \\
    250 &  0.0800 &  0.4335    &   1.46  \\
    250 &  0.1300 &  0.3587    &   1.60  \\
    250 &  0.1800 &  0.2972    &   2.20  \\
    300 &  0.0050 &  1.140     &   1.94  \\
    300 &  0.0080 &  0.9790    &   1.29  \\
    300 &  0.0130 &  0.8001    &   1.31  \\
    300 &  0.0200 &  0.7169    &   1.44  \\
    300 &  0.0320 &  0.5788    &   1.51  \\
    300 &  0.0500 &  0.4936    &   1.66  \\
    300 &  0.0800 &  0.4384    &   1.71  \\
    300 &  0.1300 &  0.3724    &   1.75  \\
    300 &  0.1800 &  0.3087    &   2.26  \\
    300 &  0.4000 &  0.1476    &   2.95  \\
    400 &  0.0080 &  0.9859    &   1.60  \\
    400 &  0.0130 &  0.8665    &   1.49  \\
    400 &  0.0200 &  0.7125    &   1.57  \\
    400 &  0.0320 &  0.5910    &   1.68  \\
    400 &  0.0500 &  0.4989    &   1.85  \\
    400 &  0.0800 &  0.4340    &   2.00  \\
    400 &  0.1300 &  0.3538    &   1.94  \\
    400 &  0.1800 &  0.3076    &   2.51  \\
    400 &  0.4000 &  0.1435    &   3.13  \\
\hline
\\
\\
\\
\\
\\
    \end{tabular}
  \end{minipage}
  \hskip.02\textwidth
  \begin{minipage}{.31\textwidth}
    \begin{tabular}{rclr}
\hline
      \Qsq &  $x$ & $\sigma_{\rm red}$ & $\delta_{\rm stat}$ \\
$[{\GeVsq}]$ &   &  &  $[\%]$ \\
\hline
    500 &  0.0080 &  0.9862    &   2.69  \\
    500 &  0.0130 &  0.8622    &   1.85  \\
    500 &  0.0200 &  0.7448    &   1.89  \\
    500 &  0.0320 &  0.6130    &   1.95  \\
    500 &  0.0500 &  0.5351    &   2.06  \\
    500 &  0.0800 &  0.4512    &   2.21  \\
    500 &  0.1300 &  0.3739    &   2.49  \\
    500 &  0.1800 &  0.3124    &   2.91  \\
    500 &  0.2500 &  0.2508    &   3.60  \\
    650 &  0.0130 &  0.8444    &   2.14  \\
    650 &  0.0200 &  0.7301    &   2.21  \\
    650 &  0.0320 &  0.6681    &   2.28  \\
    650 &  0.0500 &  0.5319    &   2.38  \\
    650 &  0.0800 &  0.4372    &   2.68  \\
    650 &  0.1300 &  0.3882    &   3.20  \\
    650 &  0.1800 &  0.3478    &   3.07  \\
    650 &  0.2500 &  0.2389    &   3.85  \\
    650 &  0.4000 &  0.1352    &   5.35  \\
    800 &  0.0130 &  0.8458    &   3.47  \\
    800 &  0.0200 &  0.7083    &   2.52  \\
    800 &  0.0320 &  0.6392    &   2.60  \\
    800 &  0.0500 &  0.5330    &   2.90  \\
    800 &  0.0800 &  0.4504    &   3.06  \\
    800 &  0.1300 &  0.3663    &   3.54  \\
    800 &  0.1800 &  0.3316    &   3.68  \\
    800 &  0.2500 &  0.2574    &   4.23  \\
    800 &  0.4000 &  0.1215    &   6.45  \\
   1000 &  0.0130 &  0.7831    &   3.64  \\
   1000 &  0.0200 &  0.7302    &   2.97  \\
   1000 &  0.0320 &  0.6470    &   2.86  \\
   1000 &  0.0500 &  0.5420    &   3.09  \\
   1000 &  0.0800 &  0.4554    &   3.40  \\
   1000 &  0.1300 &  0.3484    &   4.49  \\
   1000 &  0.1800 &  0.3044    &   4.35  \\
   1000 &  0.2500 &  0.2559    &   4.74  \\
   1000 &  0.4000 &  0.1382    &   8.52  \\
   1200 &  0.0130 &  0.8759    &   5.78  \\
   1200 &  0.0200 &  0.7496    &   3.64  \\
   1200 &  0.0320 &  0.5929    &   3.51  \\
   1200 &  0.0500 &  0.5162    &   3.52  \\
   1200 &  0.0800 &  0.4456    &   3.76  \\
   1200 &  0.1300 &  0.3656    &   4.55  \\
   1200 &  0.1800 &  0.3449    &   5.25  \\
   1200 &  0.2500 &  0.2404    &   5.97  \\
   1200 &  0.4000 &  0.1103    &   7.57  \\
\hline
\\
\\
\\
\\
    \end{tabular}
  \end{minipage}
  \hskip.02\textwidth
  \begin{minipage}{.31\textwidth}
    \begin{tabular}{rclr}
\hline
      \Qsq &  $x$ & $\sigma_{\rm red}$ & $\delta_{\rm stat}$ \\
$[{\GeVsq}]$ &   &  &  $[\%]$ \\
\hline
   1500 &  0.0200 &  0.7066    &   4.63  \\
   1500 &  0.0320 &  0.6057    &   4.29  \\
   1500 &  0.0500 &  0.5409    &   4.00  \\
   1500 &  0.0800 &  0.4435    &   4.31  \\
   1500 &  0.1300 &  0.3634    &   5.16  \\
   1500 &  0.1800 &  0.3161    &   5.38  \\
   1500 &  0.2500 &  0.2148    &   6.22  \\
   1500 &  0.4000 &  0.1278    &   7.55  \\
   1500 &  0.6500 &  0.01479   &  14.78  \\
   2000 &  0.0219 &  0.7342    &   7.48  \\
   2000 &  0.0320 &  0.5603    &   5.24  \\
   2000 &  0.0500 &  0.5596    &   4.83  \\
   2000 &  0.0800 &  0.4293    &   5.03  \\
   2000 &  0.1300 &  0.3821    &   6.71  \\
   2000 &  0.1800 &  0.3152    &   6.34  \\
   2000 &  0.2500 &  0.2608    &   6.45  \\
   2000 &  0.4000 &  0.1368    &   8.15  \\
   2000 &  0.6500 &  0.01480   &  17.19  \\
   3000 &  0.0320 &  0.6145    &   5.01  \\
   3000 &  0.0500 &  0.5424    &   4.22  \\
   3000 &  0.0800 &  0.4717    &   4.45  \\
   3000 &  0.1300 &  0.3559    &   5.53  \\
   3000 &  0.1800 &  0.3364    &   7.17  \\
   3000 &  0.2500 &  0.2359    &   6.20  \\
   3000 &  0.4000 &  0.1200    &   7.64  \\
   3000 &  0.6500 &  0.01293   &  15.66  \\
   5000 &  0.0547 &  0.5109    &   6.66  \\
   5000 &  0.0800 &  0.4688    &   4.98  \\
   5000 &  0.1300 &  0.3724    &   5.99  \\
   5000 &  0.1800 &  0.3302    &   6.59  \\
   5000 &  0.2500 &  0.2143    &   8.44  \\
   5000 &  0.4000 &  0.1151    &   9.78  \\
   5000 &  0.6500 &  0.01243   &  18.62  \\
   8000 &  0.0875 &  0.4324    &  10.53  \\
   8000 &  0.1300 &  0.3196    &   9.01  \\
   8000 &  0.1800 &  0.2936    &   9.01  \\
   8000 &  0.2500 &  0.2262    &  13.15  \\
   8000 &  0.4000 &  0.1021    &  13.63  \\
   8000 &  0.6500 &  0.01562   &  19.28  \\
  12000 &  0.1300 &  0.2127    &  27.73  \\
  12000 &  0.1800 &  0.2220    &  15.03  \\
  12000 &  0.2500 &  0.1707    &  15.33  \\
  12000 &  0.4000 &  0.1257    &  16.94  \\
  12000 &  0.6500 &  0.02261   &  24.29  \\
  20000 &  0.2500 &  0.1423    &  25.06  \\
  20000 &  0.4000 &  0.1118    &  23.67  \\
  20000 &  0.6500 &  0.006952  &  71.00  \\
  30000 &  0.4000 &  0.07828   &  51.28  \\
  30000 &  0.6500 &  0.01392   &  70.94  \\
   \hline
    \end{tabular}
  \end{minipage}
} 
  \caption{
     The NC $e^+p$ reduced cross section $\sigma_{\rm red}$ with lepton beam
     polarisation $P_e=32.5\%$ with their 
     statistical ($\delta_{\rm stat}$) uncertainties.
     The full uncertainties are available in
     ref.~\cite{Aaron:2012qi}, while the respective cross section values
     are updated according to section~\ref{sec:data} and ref.~\cite{Aaron:2012kn}.
   }
  \label{tab:NC4}
\end{table}

\begin{table}[ht]{}
  \scriptsize
  \begin{minipage}{.47\textwidth}
  \centering
   \begin{tabular}{rclr}
   \hline
\Qsq &  $x$ & $\sigma$ & $\delta_{\rm stat}$ \\ 
  $[{\GeVsq}]$ &  & $[{\rm pb/\GeVsq}]$ & $[\%]$ \\ 
\hline
    300 &   0.008 &  $2.03$        &   40.6  \\
    300 &   0.013 &  $0.934$       &   14.4  \\
    300 &   0.032 &  $0.309$       &   14.0  \\
    300 &   0.080 &  $0.785 \cdot 10^{-1}$  &   13.5  \\
    500 &   0.013 &  $0.799$       &    9.8  \\
    500 &   0.032 &  $0.252$       &    8.1  \\
    500 &   0.080 &  $0.627 \cdot 10^{-1}$  &    9.3  \\
    500 &   0.130 &  $0.348 \cdot 10^{-1}$  &   21.4  \\
   1000 &   0.013 &  $0.482$       &   10.2  \\
   1000 &   0.032 &  $0.232$       &    6.2  \\
   1000 &   0.080 &  $0.716 \cdot 10^{-1}$  &    6.4  \\
   1000 &   0.130 &  $0.339 \cdot 10^{-1}$  &   10.9  \\
   2000 &   0.032 &  $0.150$       &    5.8  \\
   2000 &   0.080 &  $0.579 \cdot 10^{-1}$  &    5.2  \\
   2000 &   0.130 &  $0.293 \cdot 10^{-1}$  &    7.4  \\
   2000 &   0.250 &  $0.106 \cdot 10^{-1}$  &   14.6  \\
   3000 &   0.080 &  $0.402 \cdot 10^{-1}$  &    5.2  \\
   3000 &   0.130 &  $0.236 \cdot 10^{-1}$  &    6.1  \\
   3000 &   0.250 &  $0.865 \cdot 10^{-2}$  &    9.2  \\
   5000 &   0.080 &  $0.265 \cdot 10^{-1}$  &    6.6  \\
   5000 &   0.130 &  $0.158 \cdot 10^{-1}$  &    5.8  \\
   5000 &   0.250 &  $0.609 \cdot 10^{-2}$  &    7.1  \\
   5000 &   0.400 &  $0.185 \cdot 10^{-2}$  &   19.2  \\
   8000 &   0.130 &  $0.107 \cdot 10^{-1}$  &    7.1  \\
   8000 &   0.250 &  $0.325 \cdot 10^{-2}$  &    7.3  \\
   8000 &   0.400 &  $0.132 \cdot 10^{-2}$  &   13.4  \\
  15000 &   0.250 &  $0.197 \cdot 10^{-2}$  &    8.4  \\
  15000 &   0.400 &  $0.492 \cdot 10^{-3}$  &   11.1  \\
  30000 &   0.400 &  $0.208 \cdot 10^{-3}$  &   17.2  \\

   \hline
   \end{tabular}
\end{minipage}
  \hskip.06\textwidth
  \begin{minipage}{.47\textwidth}
  \centering
   \begin{tabular}{rclr}
   \hline
\Qsq &  $x$ & $\sigma$ & $\delta_{\rm stat}$ \\ 
  $[{\GeVsq}]$ &  & $[{\rm pb/\GeVsq}]$ & $[\%]$ \\ 
\hline
    300 &   0.008 &  $1.18$        &   47.2  \\
    300 &   0.013 &  $0.428$       &   35.0  \\
    300 &   0.032 &  $0.129$       &   24.9  \\
    300 &   0.080 &  $0.473 \cdot 10^{-1}$  &   25.2  \\
    500 &   0.013 &  $0.412$       &   20.5  \\
    500 &   0.032 &  $0.143$       &   16.3  \\
    500 &   0.080 &  $0.368 \cdot 10^{-1}$  &   18.4  \\
    500 &   0.130 &  $0.133 \cdot 10^{-1}$  &   50.5  \\
   1000 &   0.013 &  $0.286$       &   19.9  \\
   1000 &   0.032 &  $0.116$       &   12.8  \\
   1000 &   0.080 &  $0.446 \cdot 10^{-1}$  &   12.0  \\
   1000 &   0.130 &  $0.129 \cdot 10^{-1}$  &   26.3  \\
   2000 &   0.032 &  $0.717 \cdot 10^{-1}$  &   12.7  \\
   2000 &   0.080 &  $0.234 \cdot 10^{-1}$  &   12.4  \\
   2000 &   0.130 &  $0.130 \cdot 10^{-1}$  &   16.3  \\
   2000 &   0.250 &  $0.548 \cdot 10^{-2}$  &   28.9  \\
   3000 &   0.080 &  $0.229 \cdot 10^{-1}$  &   10.1  \\
   3000 &   0.130 &  $0.923 \cdot 10^{-2}$  &   14.3  \\
   3000 &   0.250 &  $0.384 \cdot 10^{-2}$  &   19.6  \\
   5000 &   0.080 &  $0.137 \cdot 10^{-1}$  &   13.6  \\
   5000 &   0.130 &  $0.850 \cdot 10^{-2}$  &   11.7  \\
   5000 &   0.250 &  $0.283 \cdot 10^{-2}$  &   15.1  \\
   5000 &   0.400 &  $0.837 \cdot 10^{-3}$  &   40.9  \\
   8000 &   0.130 &  $0.550 \cdot 10^{-2}$  &   14.3  \\
   8000 &   0.250 &  $0.170 \cdot 10^{-2}$  &   15.4  \\
   8000 &   0.400 &  $0.514 \cdot 10^{-3}$  &   31.7  \\
  15000 &   0.250 &  $0.103 \cdot 10^{-2}$  &   17.7  \\
  15000 &   0.400 &  $0.271 \cdot 10^{-3}$  &   23.6  \\
 
   \hline
   \\ 
   \end{tabular}
  \end{minipage}
  \caption {
  The CC $e^-p$ cross section $\sigma={d^2\sigma^{\rm CC}}/{dxd\Qsq}$ for lepton polarisation
  $P_e=-25.8\%$  (left) and $P_e=36.0\%$ (right)
  with their statistical
     ($\delta_{\rm stat}$) uncertainties.
     The full uncertainties are available in
     ref.~\cite{Aaron:2012qi}, while the respective cross section values
     are updated according to section~\ref{sec:data} and ref.~\cite{Aaron:2012kn}.
   }
   \label{tab:CC1}
\end{table}

\begin{table}[ht]{}
  \scriptsize
  \begin{minipage}{.47\textwidth}
    \centering
    \begin{tabular}{rclr}
      \hline
\Qsq &  $x$ & $\sigma$ & $\delta_{\rm stat}$ \\ 
  $[{\GeVsq}]$ &  & $[{\rm pb/\GeVsq}]$ & $[\%]$ \\ 
\hline
    300 &   0.008 &  $1.21$        &   38.5  \\
    300 &   0.013 &  $0.414$       &   28.4  \\
    300 &   0.032 &  $0.102$       &   23.6  \\
    300 &   0.080 &  $0.258 \cdot 10^{-1}$  &   27.5  \\
    500 &   0.013 &  $0.286$       &   20.4  \\
    500 &   0.032 &  $0.105$       &   15.2  \\
    500 &   0.080 &  $0.386 \cdot 10^{-1}$  &   14.2  \\
    500 &   0.130 &  $0.122 \cdot 10^{-1}$  &   41.5  \\
   1000 &   0.013 &  $0.241$       &   18.4  \\
   1000 &   0.032 &  $0.124$       &    9.9  \\
   1000 &   0.080 &  $0.204 \cdot 10^{-1}$  &   13.9  \\
   1000 &   0.130 &  $0.736 \cdot 10^{-2}$  &   26.1  \\
   2000 &   0.032 &  $0.537 \cdot 10^{-1}$  &   11.3  \\
   2000 &   0.080 &  $0.157 \cdot 10^{-1}$  &   11.5  \\
   2000 &   0.130 &  $0.698 \cdot 10^{-2}$  &   17.1  \\
   2000 &   0.250 &  $0.229 \cdot 10^{-2}$  &   31.8  \\
   3000 &   0.080 &  $0.119 \cdot 10^{-1}$  &   11.3  \\
   3000 &   0.130 &  $0.544 \cdot 10^{-2}$  &   14.9  \\
   3000 &   0.250 &  $0.158 \cdot 10^{-2}$  &   23.1  \\
   5000 &   0.080 &  $0.364 \cdot 10^{-2}$  &   21.0  \\
   5000 &   0.130 &  $0.309 \cdot 10^{-2}$  &   15.6  \\
   5000 &   0.250 &  $0.816 \cdot 10^{-3}$  &   22.5  \\
   5000 &   0.400 &  $0.529 \cdot 10^{-3}$  &   40.9  \\
   8000 &   0.130 &  $0.696 \cdot 10^{-3}$  &   29.7  \\
   8000 &   0.250 &  $0.621 \cdot 10^{-3}$  &   20.5  \\
   8000 &   0.400 &  $0.802 \cdot 10^{-4}$  &   58.3  \\
  15000 &   0.250 &  $0.741 \cdot 10^{-4}$  &   46.0  \\
  15000 &   0.400 &  $0.318 \cdot 10^{-4}$  &   45.0  \\

      \hline
      \\ 
    \end{tabular}
  \end{minipage}
  \hskip.06\textwidth
  \begin{minipage}{.47\textwidth}
  \centering
   \begin{tabular}{rclr}
   \hline
\Qsq &  $x$ & $\sigma$ & $\delta_{\rm stat}$ \\ 
  $[{\GeVsq}]$ &  & $[{\rm pb/\GeVsq}]$ & $[\%]$ \\ 
\hline
    300 &   0.008 &  $0.778$       &   49.3  \\
    300 &   0.013 &  $0.593$       &   20.4  \\
    300 &   0.032 &  $0.273$       &   11.9  \\
    300 &   0.080 &  $0.519 \cdot 10^{-1}$  &   16.8  \\
    500 &   0.008 &  $1.57$        &   23.2  \\
    500 &   0.013 &  $0.670$       &   11.4  \\
    500 &   0.032 &  $0.252$       &    8.5  \\
    500 &   0.080 &  $0.603 \cdot 10^{-1}$  &    9.8  \\
    500 &   0.130 &  $0.268 \cdot 10^{-1}$  &   23.7  \\
   1000 &   0.013 &  $0.392$       &   12.5  \\
   1000 &   0.032 &  $0.176$       &    7.4  \\
   1000 &   0.080 &  $0.512 \cdot 10^{-1}$  &    7.8  \\
   1000 &   0.130 &  $0.267 \cdot 10^{-1}$  &   12.1  \\
   2000 &   0.032 &  $0.104$       &    7.3  \\
   2000 &   0.080 &  $0.371 \cdot 10^{-1}$  &    6.6  \\
   2000 &   0.130 &  $0.165 \cdot 10^{-1}$  &    9.9  \\
   2000 &   0.250 &  $0.473 \cdot 10^{-2}$  &   19.3  \\
   3000 &   0.080 &  $0.247 \cdot 10^{-1}$  &    6.9  \\
   3000 &   0.130 &  $0.154 \cdot 10^{-1}$  &    8.0  \\
   3000 &   0.250 &  $0.260 \cdot 10^{-2}$  &   15.9  \\
   5000 &   0.080 &  $1.000 \cdot 10^{-2}$  &   10.9  \\
   5000 &   0.130 &  $0.636 \cdot 10^{-2}$  &    9.8  \\
   5000 &   0.250 &  $0.187 \cdot 10^{-2}$  &   13.3  \\
   5000 &   0.400 &  $0.873 \cdot 10^{-3}$  &   25.0  \\
   8000 &   0.130 &  $0.217 \cdot 10^{-2}$  &   15.4  \\
   8000 &   0.250 &  $0.100 \cdot 10^{-2}$  &   13.5  \\
   8000 &   0.400 &  $0.299 \cdot 10^{-3}$  &   27.8  \\
  15000 &   0.250 &  $0.315 \cdot 10^{-3}$  &   19.0  \\
  15000 &   0.400 &  $0.370 \cdot 10^{-4}$  &   38.0  \\

   \hline
   \end{tabular}
  \end{minipage}
\caption{
  The CC $e^+p$ cross section $\sigma={d^2\sigma^{\rm CC}}/{dxd\Qsq}$ for lepton polarisation
     $P_e=-37.0\%$ (left) and $P_e=32.5\%$ (right) with their statistical
     ($\delta_{\rm stat}$) uncertainties.
     The full uncertainties are available in
     ref.~\cite{Aaron:2012qi}, while the respective cross section values
     are updated according to section~\ref{sec:data} and ref.~\cite{Aaron:2012kn}.
   }
  \label{tab:CC2}
\end{table}

\clearpage
\section{Results of fits with many parameters}
\label{appx:results}
\renewcommand{\arraystretch}{1.25} 
Table~\ref{tab:rhopwithcorrelations} quotes the fit of
\rhop{}, \kapp{} and \rhopW{} parameters and their correlation
coefficients.
Tables~\ref{tab:rhopkapq2:q} to~\ref{tab:rhopW:f%
} quote fits
of scale-dependent \rhop{}, \kapp{} and \rhopW{} parameters and their
correlation coefficients.

\begin{table}[ht]
  \footnotesize
  \centering
   \begin{tabular}{lr@{$\,=\,$}c@{$\,\pm\,$}l|cccc} 
   \hline
   Fit parameters & \multicolumn{3}{c}{Result} & \multicolumn{4}{c}{Correlation} \\
   \hline
   \rhopu+\kappu+\rhopd+\kappd+PDF
   & \rhopu & 1.53 & 0.35 & 1.00 \\                      
   & \kappu & 1.26 & 0.14 & 0.29  & 1.00 \\               
   & \rhopd & 0.18 & 0.39 &$-0.86$& $-0.26$ & 1.00 \\        
   & \kappd &$-6.4$&10.5  &$-0.84$& $-0.34$ & 0.993 & 1.00 \\
   \hline
   \rhop{,q}+\kapp{,q}+\rhop{,e}+\kapp{,e}+PDF
   & \rhop{,q} & 1.99 & 1.91 & 1.00 \\        
   & \kapp{,q} & 0.93 & 0.12 & $-0.02$ & 1.00 \\
   & \rhop{,e} & 0.59 & 0.58 & $-0.99$ & 0.09   & 1.00 \\       
   & \kapp{,e} & 0.98 & 0.06 & $-0.25$ &$-0.10$ & 0.33  & 1.00 \\
   \hline
   \rhop{,f}$+$\kapp{,f}$+$\rhopW{,f}+PDF
   & \rhop{f}   & 1.09  & 0.07   & 1.00 & \\
   & \kapp{f}   & 0.97  & 0.05   & 0.82 & 1.00  \\ %
   & \rhopW{,f} & 1.004 & 0.008  & 0.03 & $-0.12$ & 1.00\\
   \hline
   \end{tabular}
   \caption{
     Results for $\rhop{}$, $\kapp{}$ and \rhopW{} parameters, and their
     correlation coefficients, from
     fits with more than two EW parameters.
     For the \rhopd+\kappd+\rhopu+\kappu+PDF fit
     the uncertainties are only approximate since \chisq\ is not
     described by a quadratic dependence  on the fit parameters. 
     The uncertainties quoted correspond to the total uncertainties.
   }
   \label{tab:rhopwithcorrelations}
\end{table}

\begin{table}[ht]
  \footnotesize
  \centering
   \begin{tabular}{ccr@{$\,\pm\,$}l|cccccccc} 
   \hline
   \Qsq\ range $[\GeVsq]$ & Parameter & \multicolumn{2}{c}{Result} & \multicolumn{8}{c}{Correlation} \\
   \hline
     $[561,1778]$    & \rhop{,q} &   2.05 &  0.50 &  1.00  \\
     $[1778,6000]$   & \rhop{,q} &   1.06 &  0.16 &  0.11 & 1.00  \\
     $[6000,16680]$  & \rhop{,q} &   1.17 &  0.18 &  0.05 & 0.14 & 1.00 \\
     $[16680,77000]$ & \rhop{,q} &   1.59 &  0.42 &  0.01 & 0.07 & 0.11 & 1.00 \\
     $[561,1778]$    & \kapp{,q} &   1.21 &  0.15 &  0.75 & 0.07 & 0.03 &  0.01 & 1.00  \\
     $[1778,6000]$   & \kapp{,q} &   0.92 &  0.16 &  0.05 & 0.72 & 0.10 &  0.05 & 0.07 & 1.00 \\
     $[6000,16680]$  & \kapp{,q} &   1.02 &  0.19 &  0.02 & 0.09 & 0.63 &  0.07 & 0.03 & 0.09 & 1.00  \\
     $[16680,77000]$ & \kapp{,q} &   0.41 &  0.33 &  0.00 & 0.07 & 0.11 &  0.70 & 0.01 & 0.06 & 0.09 &  1.00 \\
   \hline
   \end{tabular}
   \caption{
     Results for the \rhop{,q} and \kapp{,q} parameters determined at different values of \Qsq.
     The \Qsq\ range of the data selection, and the correlation
     coefficients of the fit parameters are indicated.
   }
   \label{tab:rhopkapq2:q}
\end{table}

\begin{table}[ht]
  \footnotesize
  \centering
   \begin{tabular}{ccr@{$\,\pm\,$}l|cccccccc} 
   \hline
   \Qsq\ range $[\GeVsq]$ & Parameter & \multicolumn{2}{c}{Result} & \multicolumn{8}{c}{Correlation} \\
   \hline
   $[561,1778]$    & \rhop{,e} &  1.51  &  0.34  &  1.00  \\                    
   $[1778,6000]$   & \rhop{,e} &  1.10  &  0.18  &  0.06 & 1.00  \\             
   $[6000,16680]$  & \rhop{,e} &  1.14  &  0.24  &  0.03 & 0.17 & 1.00  \\      
   $[16680,77000]$ & \rhop{,e} &  1.19  &  0.34  &  0.02 & 0.11 & 0.16 & 1.00 \\
   $[561,1778]$    & \kapp{,e} &  0.99  &  0.06  &  0.66 & 0.07 & 0.04 & 0.03 & 1.00 \\                      
   $[1778,6000]$   & \kapp{,e} &  0.99  &  0.07  &  0.05 & 0.76 & 0.15 & 0.09 & 0.16 & 1.00  \\              
   $[6000,16680]$  & \kapp{,e} &  0.98  &  0.13  &  0.02 & 0.15 & 0.83 & 0.13 & 0.07 & 0.16 & 1.00  \\       
   $[16680,77000]$ & \kapp{,e} &  0.53  &  0.21  &  0.01 & 0.10 & 0.13 & 0.72 & 0.03 & 0.10 & 0.13 & 1.00  \\
   \hline
   \end{tabular}
   \caption{
     Results for the \rhop{,e} and \kapp{,e} parameters determined at different values of \Qsq.
     The \Qsq\ range of the data selection, and the correlation
     coefficients of the fit parameters are indicated.
   }
     \label{tab:rhopkapq2:e}
\end{table}

\begin{table}[ht]
  \footnotesize
  \centering
   \begin{tabular}{ccr@{$\,\pm\,$}l|cccccccc} 
   \hline
   \Qsq\ range $[\GeVsq]$ & Parameter & \multicolumn{2}{c}{Result} & \multicolumn{8}{c}{Correlation} \\
   \hline
   $[561,1778]$    & \rhop{,f} &  1.28 &  0.19  &  1.00 \\
   $[1778,6000]$   & \rhop{,f} &  1.03 &  0.09  &  0.07 & 1.00 \\
   $[6000,16680]$  & \rhop{,f} &  1.07 &  0.11  &  0.03 & 0.19 & 1.00  \\
   $[16680,77000]$ & \rhop{,f} &  1.09 &  0.17  &  0.02 & 0.11 & 0.16 & 1.00 \\
   $[561,1778]$    & \kapp{,f} &  1.01 &  0.07  &  0.83 & 0.07 & 0.04 & 0.03 & 1.00 \\
   $[1778,6000]$   & \kapp{,f} &  0.98 &  0.06  &  0.07 & 0.80 & 0.17 & 0.10 & 0.14 & 1.00 \\
   $[6000,16680]$  & \kapp{,f} &  0.99 &  0.09  &  0.03 & 0.16 & 0.83 & 0.13 & 0.06 & 0.18 & 1.00 \\
   $[16680,77000]$ & \kapp{,f} &  0.69 &  0.13  &  0.02 & 0.10 & 0.14 & 0.77 & 0.04 & 0.10 & 0.13 & 1.00 \\
   \hline
   \end{tabular}
   \caption{
     Results for the \rhop{,f} and \kapp{,f} parameters determined at different values of \Qsq.
     The \Qsq\ range of the data selection, and the correlation
     coefficients of the fit parameters are indicated.
   }
     \label{tab:rhopkapq2:f}
\end{table}

\begin{table}[ht]
  \footnotesize
  \centering
   \begin{tabular}{ccr@{$\,\pm\,$}l|cccc} 
   \hline
   \Qsq\ range $[\GeVsq]$ & Parameter & \multicolumn{2}{c}{Result} & \multicolumn{4}{c}{Correlation} \\
   \hline
   $[224,708]$    & \rhopW{,eq} &  0.948 &  0.030 &  1.00  \\
   $[708,2239]$   & \rhopW{,eq} &  0.993 &  0.014 &  0.40 & 1.00  \\
   $[2239,7079]$  & \rhopW{,eq} &  0.993 &  0.013 &  0.15 & 0.17 & 1.00  \\
   $[7079,25119]$ & \rhopW{,eq} &  1.008 &  0.020 &$-0.03$& 0.01 & 0.12 & 1.00 \\
   \hline
   \end{tabular}
   \caption{
     Results for the \rhopW{,eq} parameters determined at different values of \Qsq.
     The \Qsq\ range of the data selection, and the correlation
     coefficients of the fit parameters are indicated.
   }
     \label{tab:rhopW:q}
\end{table}

\begin{table}[ht]
  \footnotesize
  \centering
   \begin{tabular}{ccr@{$\,\pm\,$}l|cccc} 
   \hline
   \Qsq\ range $[\GeVsq]$ & Parameter & \multicolumn{2}{c}{Result} & \multicolumn{4}{c}{Correlation} \\
   \hline
   $[224,708]$    & \rhopW{,e\bar{q}} &    1.018 &  0.045 &  1.00  \\
   $[708,2239]$   & \rhopW{,e\bar{q}} &    1.054 &  0.041 &  0.63 & 1.00  \\
   $[2239,7079]$  & \rhopW{,e\bar{q}} &    1.062 &  0.046 &  0.48 & 0.67 & 1.00  \\
   $[7079,25119]$ & \rhopW{,e\bar{q}} &    1.010 &  0.075 &  0.14 & 0.29 & 0.50 & 1.00 \\
   \hline
   \end{tabular}
   \caption{
     Results for the \rhopW{,e\bar{q}} parameters determined at different values of \Qsq.
     The \Qsq\ range of the data selection, and the correlation
     coefficients of the fit parameters are indicated.
   }
     \label{tab:rhopW:qb}
\end{table}

\begin{table}[ht]
  \footnotesize
  \centering
   \begin{tabular}{ccr@{$\,\pm\,$}l|cccc} 
   \hline
   \Qsq\ range $[\GeVsq]$ & Parameter & \multicolumn{2}{c}{Result} & \multicolumn{4}{c}{Correlation} \\
   \hline
   $[224,708]$    & \rhopW{,f} &    0.976 &  0.018 &  1.00  \\
   $[708,2239]$   & \rhopW{,f} &    0.998 &  0.011 &  0.47 &  1.00  \\
   $[2239,7079]$  & \rhopW{,f} &    0.999 &  0.010 &  0.19 &  0.24 & 1.00  \\
   $[7079,25119]$ & \rhopW{,f} &    1.004 &  0.017 &$-0.06$&$-0.01$& 0.12 & 1.00 \\
   \hline
   \end{tabular}
   \caption{
     Results for the \rhopW{,f} parameters determined at different values of \Qsq.
     The \Qsq\ range of the data selection, and the correlation
     coefficients of the fit parameters are indicated.
   }
     \label{tab:rhopW:f}
\end{table}


\clearpage

\begin{flushleft}
\bibliography{desy18-080}
\end{flushleft}
\end{document}